# Instability Caused by Integration of IBRs under Strong Grid Connections — A Practical Case Study on Large-scale Energy Storage Systems

Qiang Fu, *Member, IEEE*, Siqi Bu, *Senior Member, IEEE*, Zijun Bin, Peng Li, and Tong Wang, *Senior Member, IEEE*

***Abstract*— Inverter-based resources (IBRs) with grid-following (GFL) control have been known to cause converter-driven stability issues under weak grid conditions. However, this study investigates a practical scenario involving large-scale energy storage systems (ESSs) using GFL control under strong grid conditions. These ESSs generate a 150 Hz oscillation in the *d-q* coordinates while providing both capacitive and inductive reactive power support to a bulk power system. To preserve manufacturer confidentiality, the stability analysis is conducted by focusing on the equivalent dynamics of ESSs rather than getting involved in their internal parameters. The results show that dynamic interactions among multiple ESSs intensify as their scale increases. Although each individual ESS is designed to be stable, instability occurs from the superimposed effect of ESSs. A damping index is proposed and indicates that the superimposed effect always induces negative damping to the system stability, and its magnitude grows as the ESS scale extends. Finally, key factors are identified, mitigation strategies are proposed, and the conclusions are validated. This study provides practical insights into instabilities caused by large-scale ESSs even under strong grid conditions, and emphasizes the importance of accounting for grid-side dynamic interactions in large-scale IBRs.**

## I. Introduction

ENERGY storage systems (ESSs) are widely deployed to mitigate active power fluctuations caused by renewable energy sources (RESs) and to provide reactive power support for enhanced AC voltage controllability in power systems [1], [2]. ESSs are thus expected to be a hallmark of future electrical grids [3]. Beyond ensuring power balance [4], ESSs also offer dynamic functional support to power systems due to their high controllability. Such capabilities include reducing overvoltage during faults [5], addressing voltage excursions [6], and alleviating frequency decline during large disturbances [7], [8]. However, considering that ESSs are power electronic devices, the power conversion systems (PCSs) of grid-side converters of ESSs can also introduce converter-driven stability (CDS) issues [9], which require further attention. In this paper, we focus on the PCSs utilizing grid-following (GFL) control, where the AC power system connected to ESSs is dominated by traditional synchronous generators (SGs). In this context, ESSs offer active and reactive power support rather than grid-forming services.

The well-known outer control loops of a grid-side converter utilizing GFL control include the DC voltage, active power, and reactive power control loops. For a grid-side converter utilizing DC voltage and reactive power outer control loops, the DC-DC converter at the source side of ESSs normally controls the active power as a constant. Consequently, the conclusions drawn from the grid-side converters of RESs can also be applied to those of ESSs because their dynamic equations are the same. Instability risks frequently arise under weak grid conditions [10], including the well-documented synchronization instability of the phase-locked loop (PLL) [11], [12]. Other risks involve voltage oscillations caused by the bandwidth limitations of the inner control loop [13], AC current oscillations due to improper filter parameters [14], and strong dynamic interactions between the DC and AC components [15]. These stability mechanisms are analogous to those of grid-side converters of RESs [16].

More frequently, the grid-side converter of an ESS uses active and reactive power outer control loops since source-side energy storage is like a DC voltage source especially for a battery-based ESS. In such cases, the instability mechanism is typically attributed to the negative impedance characteristics of the grid-side converter. For instance, [17] concluded that the stability of overall power system decreases as load power increases when the grid-side converter operates as a constant power load. A method to determine the maximum permissible number and capacity of constant power loads was proposed in [18]. Additionally, under weak grid conditions, the risk of instability is further amplified. The instability risks of the PLL under weak grid conditions have been highlighted in [19], and [20] demonstrated that feedforward power loops can induce high-gain instability under such conditions.

In previous studies, the reactive power outer control loop has been identified as a factor contributing to system instability. For example, [21] indicated that under strong grid conditions, the influence of the reactive power outer control loop on the PLL-induced instability is weak, whereas it becomes significant under weak grid connections. Under the latter scenario, [22]

Qiang Fu and Siqi Bu are with the Department of Electrical and Electronic Engineering at The Hong Kong Polytechnic University. Zijun Bin and Peng Li are with the Electric Power Science Research Institute of the State Grid Jiangsu Electric Power Co., Ltd., while Tong Wang is with the State Key Laboratory of Alternate Electrical Power Systems with Renewable Energy Sources at North China Electric Power University.
This work is supported by the State Key Laboratory of Alternate Electrical Power System with Renewable Energy Sources (Grant No. LAPS24012), and the National Natural Science Foundation of China (Grant No. 52407129). We also acknowledge the support from Hong Kong Scholars Program.
Corresponding author: Siqi Bu, Email: siqi.bu@gmail.com.

reported that injecting negative reactive power can enhance system stability. Similarly, [23] proposed a strategy that prioritizes reactive power to improve both PLL and DC link stability related to ESSs. The impact of reactive power on the AC voltage stability was more pronounced. The studies in [24] and [25] indicated that different reactive power outer control parameters and power flow conditions lead to varying stability outcomes. Therefore, these findings highlight the crucial role of reactive power outer control and its flow in maintaining system stability. Weak grid connections and negative impedance under single power flow conditions are two common mechanisms that contribute to system instabilities.

However, several research gaps remain in existing studies. First, improper reactive power distribution caused by converters has been justified shown to cause static AC voltage instability, but there is, to the best of our knowledge, no practical report on dynamic instability induced by the reactive power outer control loops in grid-side converters of MW-level ESSs. Second, most instabilities identified in the literature occur under weak grid connections. Numerous studies have suggested that strong grids can enhance the CDS under GFL control. Consequently, the potential instability risks under strong grid connections receive relatively little attention compared to those under weak grid connections. Third, although the concept of negative impedance has been widely adopted to explain instability mechanisms and is effective for analyzing the instability by dividing the power system into two subsystems. This division makes it difficult to conduct an in-depth analysis within the cluster-level dynamic interactions.

The major practical contribution of this paper is that a new type of SSOs caused by large-scale ESSs is reported, which is different from traditional ones in the following aspects:

1) The length of the transmission line between the ESSs and the external power system is about 35 km, with a short-circuit ratio (SCR) of around 7, which is typical for a power system under strong grid conditions [26].

2) The large-scale centralized ESSs have a rated power of 240 MW and a rated voltage level of 220 kV. There are no other AC devices, such as series compensation devices. Therefore, the SSO is caused by the dynamic interaction among multiple ESSs rather than the poor parameter tuning of a single ESS.

3) The SSO is observed in both inductive and capacitive reactive power scenarios, indicating that it mainly depends on the amplitude, regardless of the direction of power flow, which distinguishes those from traditional constant power loads, where the impedance is negative only under a single power flow direction and becomes positive when the power flow reverses.

This paper also conducts an in-depth theoretical investigation and extends the conclusions to various network topologies of ESSs, providing the following novel insights:

1) It is theoretically verified that dynamic interactions among large-scale ESSs can be intensified due to the superimposed effect of ESSs, leading to system instability under strong grid conditions. This instability mechanism is different from those of instability caused by a single converter [27], or the dynamic interaction between two converters or subsystems [28].

2) A damping index is proposed to quantify the superimposed effect of ESSs, which is applicable to all network topologies. It shows that as the number of ESSs increases, the dynamic interaction among ESSs intensifies, and the superimposed effect provides more negative damping to the oscillation, highlighting the importance of limiting the scale of ESSs.

3) From both practical and theoretical perspectives, this study demonstrates that system instabilities can co-occur when ESSs employ functional control to provide support to the connected power systems. It highlights the necessity of functional control designs that account for system-level stability beyond converter self-limitations.

To bridge the gap between the confidentiality requirement of manufacturers and the accurate data need for academic research, a stability analysis scheme is proposed and demonstrated in practice. This not only safeguards the confidentiality of manufacturers but also retains sufficient dynamic information of the ESSs for research purposes, thereby collectively ensuring the stable operation of the power system.

The remainder of this paper is organized as follows: Section II reports the practical oscillation observed in a large-scale ESS project. Section III develops a mathematical model of the power system connected with large-scale ESSs. Section IV analyzes the mechanism of superimposed effect of ESSs, identifies major factors, and proposes oscillation mitigation methods. Section V extends findings to more general scenarios, and analyzes the errors caused by the simplifications. Section VI validates the conclusions based on the SIMULINK platform. Section VII summarizes the key insights and inspirations for other IBRs.

## II. Report on the Oscillation Observed in Practical Energy Storage Systems

### A. Practical Power System Connected with ESSs

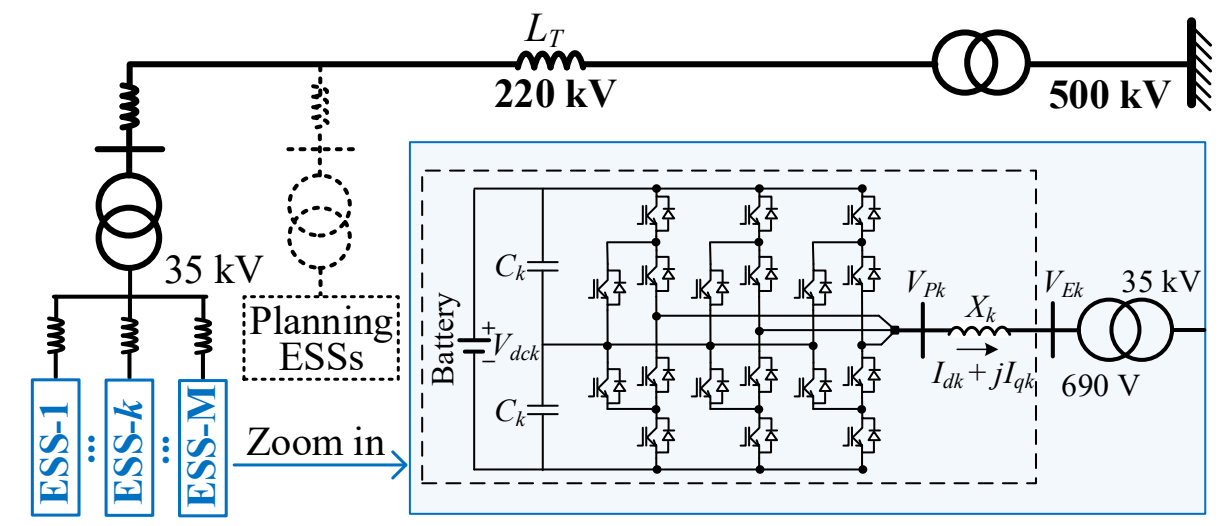

Fig. 1. Configuration of large-scale ESSs in Jiangsu Province, China.

Table I
Parameters of the practical power system

| Parameters | Values |
|---|---|
| Rated power of each ESS | 1.575 MW |
| Rated DC voltage of each ESS | ± 600 V |
| Rated AC voltage of each ESS | 690 V |
| DC side capacitance of each ESS | 36*750 uF |
| Reactance of the AC-side filter of each ESS | 100 uH |
| Total number of ESSs | 144 |
| Rated power of the power system | 240 MW |
| Rated voltage of the power system | 220 kV |
| Rated frequency of the power system | 50 Hz |
| Impedance of the AC transmission line | 28 Ω |

The practical configuration of a power system connected with large-scale ESSs is illustrated in Fig. 1. The ESSs are integrated in parallel at a common node, and then connected to the external power system through an AC transmission line (The total inductance of the transmission line is $L_T$, and the inductance per unit length is $L_0$). Each ESS consists of a three-level converter [29] for bidirectional power exchange between the ESS and the AC power system. The detailed parameters and three-level converter topology are provided in Table I and Fig. 1.

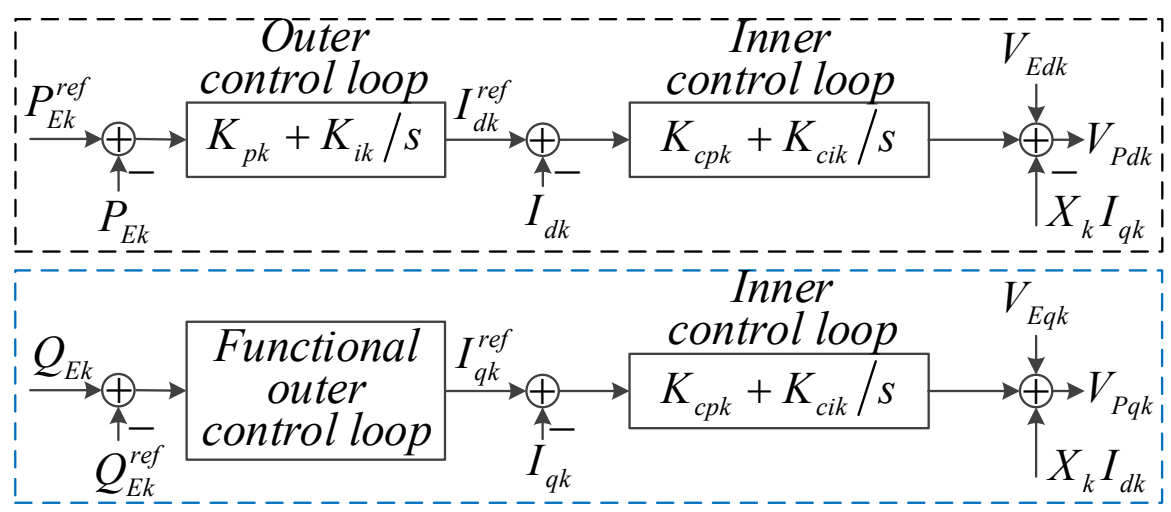

Fig. 2. Configuration of the control loops adopted by the PCS of ESSs.

The control loops of the ESS PCS are depicted in Fig. 2 [30]. $P_{Ek}$ and $Q_{Ek}$ are the active and reactive power output from the $k^{\text{th}}$ ESS, respectively. $V_{Ek}$ and $V_{Pk}$ are the AC voltages at the point of common coupling (PCC) and the output from the PCS of the $k^{\text{th}}$ ESS, respectively. $X_k$ is the reactance of the AC-side filter of the $k^{\text{th}}$ ESS. $I_k$ is the AC current output from the $k^{\text{th}}$ ESS. Subscripts $d$ and $q$ correspond to the $d$- and $q$-axis components of variables, while subscripts $x$ and $y$ correspond to $x$- and $y$-axis components of variables. The $d$-$q$ coordinates represent the local coordinates of the PCS, and the $x$-$y$ coordinates represent the global coordinates of the power system. $K_{pk} + K_{ik}s^{-1}$ and $K_{cpk} + K_{cik}s^{-1}$ are the transfer functions of the outer and inner control loops of the $k^{\text{th}}$ ESS, respectively.

Notably, the outer control loop for reactive power does not use a conventional PI controller as in the outer control loop for active power. A specialized functional controller developed by the ESS manufacturer is adopted in this outer control loop. To precisely delineate it from PI control structures and clarify its unique nature, we refer to it as the "functional outer control loop." This design allows auto-tuning of control parameters considering the operational states, making the adjustment of reactive power more sensitive as the reactive power support becomes heavier. Due to confidentiality constraints, the detailed structure of the functional outer control loop cannot be disclosed. In Section III-A, a scheme is proposed and demonstrated in practice to address this issue by matching the dynamics of the functional outer control loop around 150 Hz.

## B. Report on the Reactive Power–related Oscillation

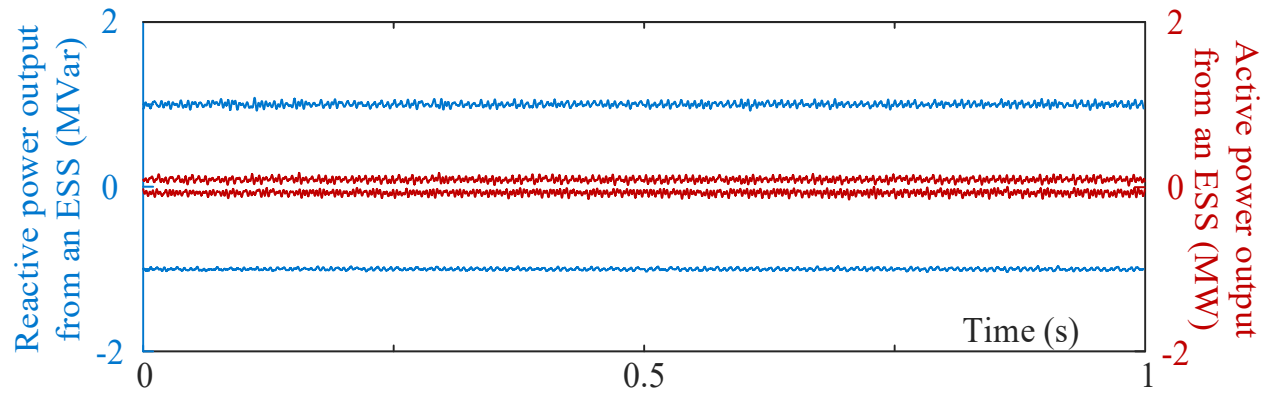

Fig. 3. Active and reactive power curves when an ESS operates under rated reactive power (with both positive and negative power flow directions)

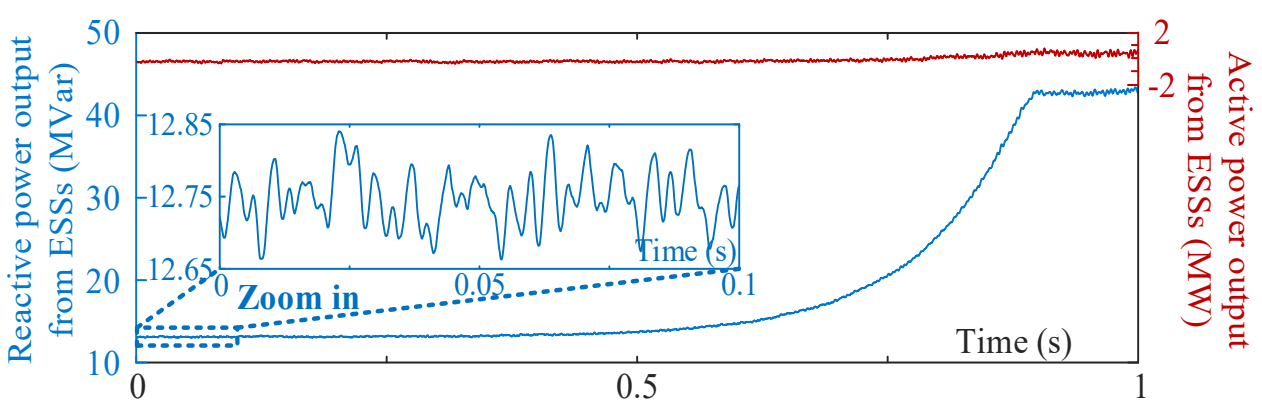

Fig. 4. Active and reactive power curves when the reactive power output from ESSs ranges from ~13 MVar to ~42 MVar (capacitive reactive power).

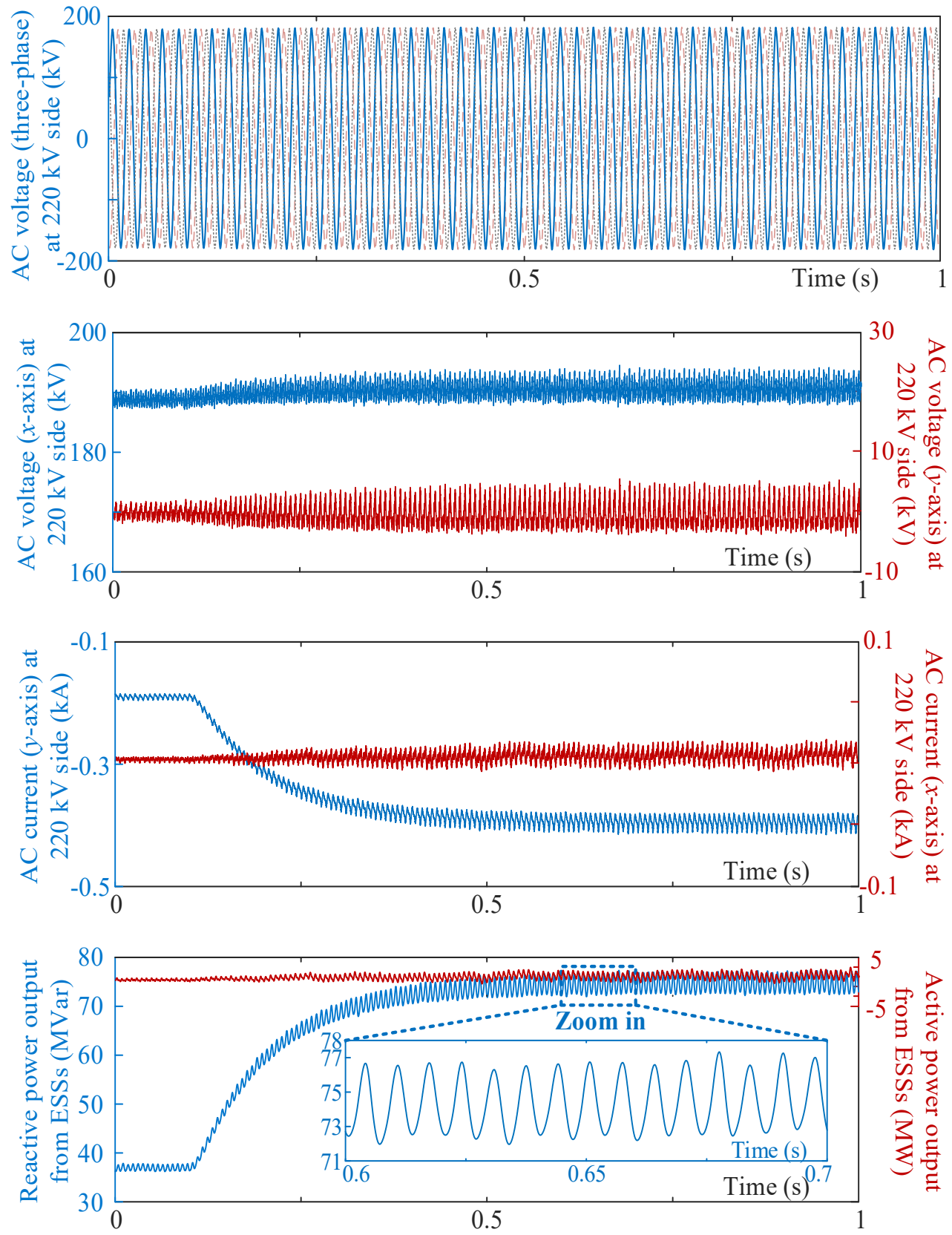

Fig. 5. Variable curves when the reactive power output from ESSs ranges from ~38 MVar to ~75 MVar (capacitive reactive power).

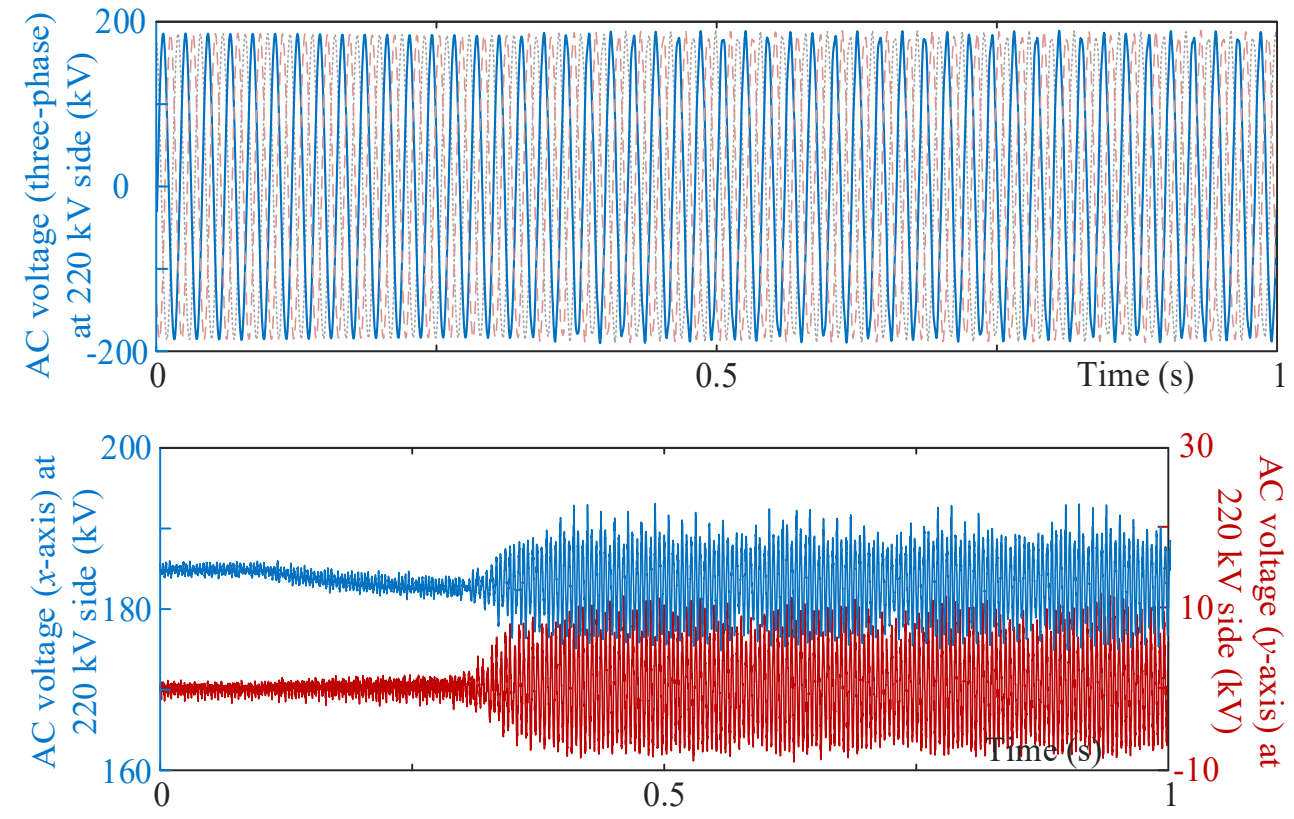

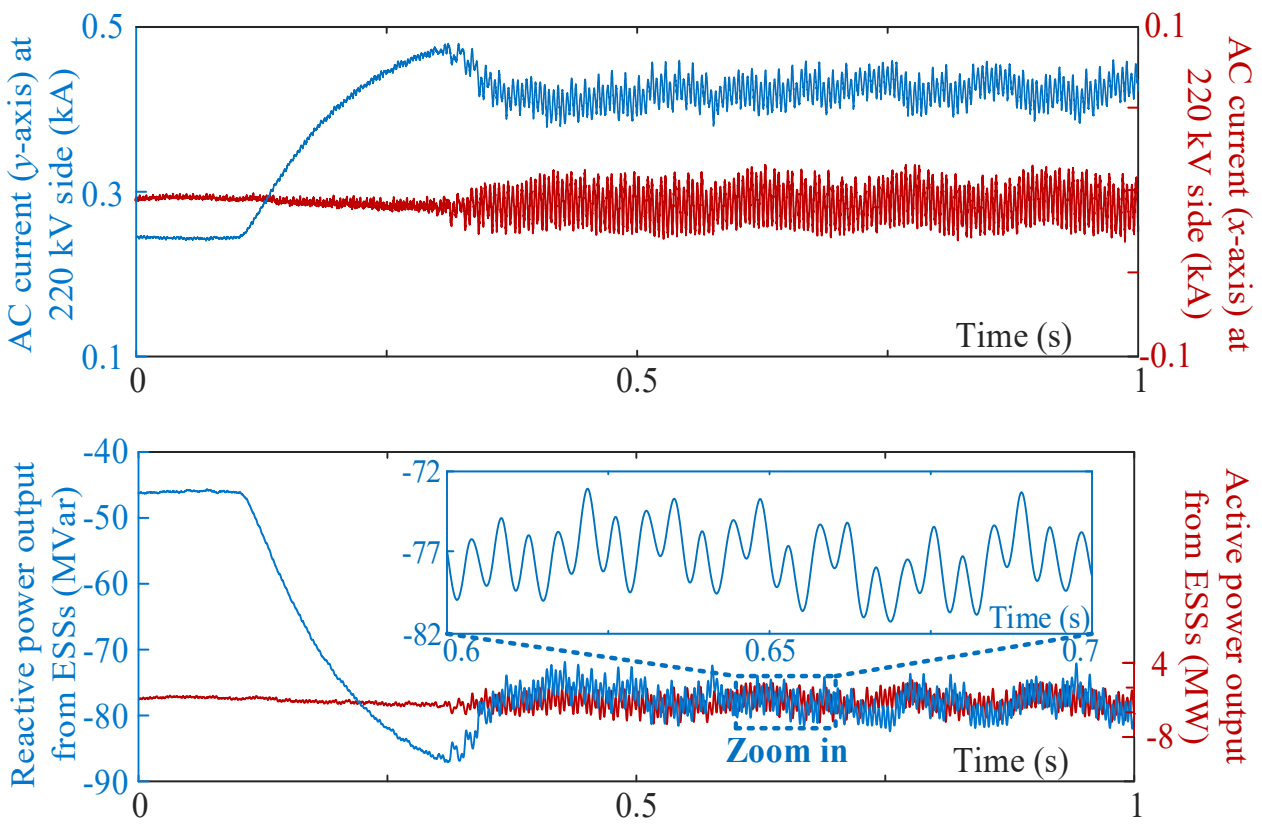

Fig. 6. Variable curves when the reactive power output from ESSs ranges from ~-47 MVar to ~-85 MVar (inductive reactive power).

As shown in Fig. 3, the single ESS operates stably under the rated reactive power with both positive and negative power flow directions. This indicates that the single ESS satisfies the requirements for the grid connection and that the instability is not caused by the ESS PCS itself.

Subsequently, two major cases are reported in this subsection: In the first case, all ESSs are connected to the grid and supply capacitive reactive power (the direction of the power flow is positive) to the external power system. The corresponding dynamic response curves are shown in Figs. 4 and 5. In the second case, all ESSs supply inductive reactive power (the direction of the power flow is negative), and the corresponding curves are shown in Fig. 6.

In Fig. 4, as the number of grid-connected ESSs increases, the amplitude of the capacitive reactive power increases from ~12 MVar to ~42 MVar. Correspondingly, the amplitude of the oscillation in the reactive power begins to increase. From Fig. 5, when the amplitude of capacitive reactive power further increases to ~75 MVar (about 0.31 p.u.), a significant oscillation in reactive power appears. Consequently, the active power becomes oscillatory, even though the amplitude of the ESS active power remains close to zero. In addition, the oscillation can be observed in the curves of other variables.

Similarly, in Fig. 6, when the inductive reactive power increases from ~47 MVar to ~80 MVar (about 0.33 p.u.), an obvious oscillation in reactive power can be observed, which also leads to oscillation in active power and other variables. Because the active power remains zero, these oscillations can be attributed to the reactive power control loop during bidirectional reactive power support to the external power system.

By comparing the results shown in Figs. 3 to 5, it can be easily concluded that the oscillation is not generated by a single ESS but rather by the increased number of ESSs. The dynamic characteristics of large-scale ESSs cannot be regarded as a multiple of those of a single ESS.

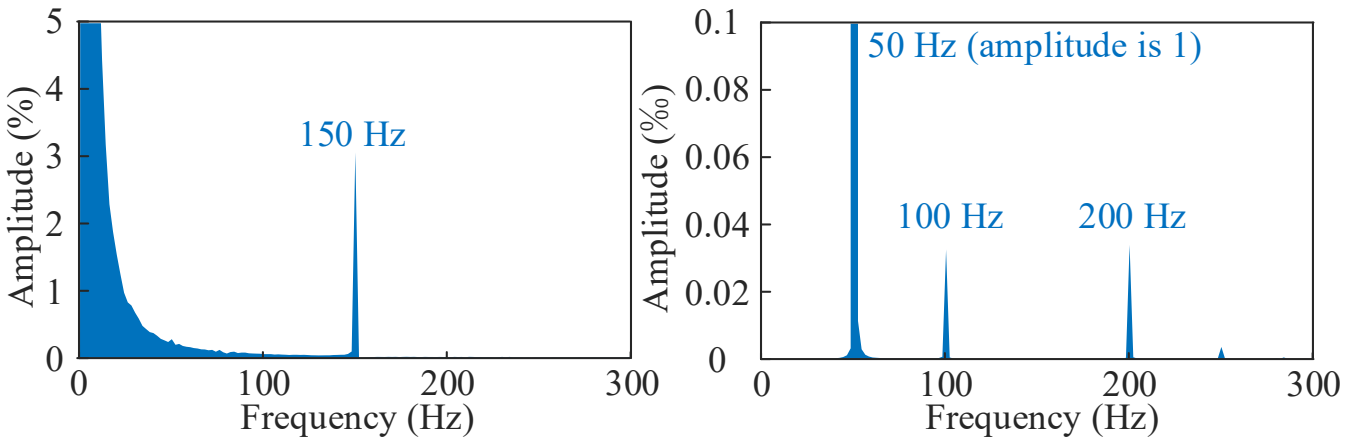

Fig. 7. FFT analysis results of reactive power (left) and three-phase AC voltage (right).

Further insights are provided by the FFT analysis in Fig. 7, which reveals an oscillation frequency of 150 Hz in reactive power (consistent in both the *d*-*q* and the *x*-*y* coordinates). The corresponding oscillation frequencies in the three-phase AC voltage are 100 Hz and 200 Hz. This phenomenon is explained by (1) to (5) [27]. Moreover, an oscillation component with a frequency of 350 Hz is observed in the three-phase coordinates, which is attributed to switching dynamics of three-level converters [31] rather than the control loops, and thus is not further considered.

Denote the oscillation frequency as $f_d$ in the *d*-*q* coordinates, and thus the mathematical expressions of $V_{Edk}$, $V_{Eqk}$, $I_{dk}$ and $I_{qk}$ can be formulated as

$$\begin{aligned} V_{Edk} &= V_{Edk,0} + \beta_{dk} \cos\left(2\pi f_d t\right) \\ V_{Eqk} &= V_{Eqk,0} + \beta_{qk} \cos\left(2\pi f_d t\right) \\ I_{dk} &= I_{dk,0} + \alpha_{dk} \cos\left(2\pi f_d t\right) \\ I_{qk} &= I_{qk,0} + \alpha_{qk} \cos\left(2\pi f_d t\right) \end{aligned} \tag{1}$$

where the subscript 0 denotes the variable's value at the steady state. The coefficients $\alpha_{dk}$, $\alpha_{qk}$, $\beta_{dk}$, and $\beta_{qk}$ denote the amplitudes of the oscillation component at $f_d$, and are significantly smaller than the steady-state values. Specifically, $\alpha_{dk} << I_{dk}$, $\alpha_{qk} << I_{qk}$, $\beta_{dk} << V_{Edk}$, and $\beta_{qk} << V_{Eqk}$.

Considering that

$$Q_{Ek} = V_{Eqk} I_{dk} - V_{Edk} I_{qk} \tag{2}$$

Substituting (1) into (2), yields

$$\begin{aligned} Q_{Ek} \approx & -V_{Edk,0} I_{qk,0} - \left(I_{qk,0}\beta_{dk} + V_{Edk,0}\alpha_{qk}\right)\cos\left(2\pi f_d t\right) \\ & + V_{Eqk,0} I_{dk,0} + \left(I_{dk,0}\beta_{qk} + V_{Eqk,0}\alpha_{dk}\right)\cos\left(2\pi f_d t\right) \end{aligned} \tag{3}$$

where high-order small items are ignored.

Therefore, as seen from (3), the reactive power will also exhibit an oscillation at frequency $f_d$ if the oscillation frequency of variables in the *d*-*q* coordinates is $f_d$.

Regarding the three-phase AC voltage, using phase A as an example, it can be obtained by

$$V_{EAk} = V_{Edk} \cos\left(2\pi f_0 t\right) + V_{Eqk} \cos\left(2\pi f_0 t\right) \tag{4}$$

Substituting (1) into (4), yields

$$
\begin{aligned}
&V_{EAk} = V_{Edk,0}\cos(2\pi f_0 t) + \beta_{dk}\cos(2\pi f_d t)\cos(2\pi f_0 t) \\
&+V_{Eqk,0}\cos(2\pi f_0 t) + \beta_{qk}\cos(2\pi f_d t)\cos(2\pi f_0 t) \\
&\qquad\qquad \Downarrow \textit{Prosthaphaeresis} \\
&V_{EAk} = V_{Edk,0}\cos(2\pi f_0 t) + V_{Eqk,0}\cos(2\pi f_0 t) \\
&+\frac{\beta_{dk}+\beta_{qk}}{2}\left(\cos(2\pi(f_0+f_d)t) + \cos(2\pi(f_0-f_d)t)\right)
\end{aligned} \tag{5}
$$

From (5), it is evident that when the frequency in the *d*-*q* coordinates is $f_d$, the oscillation frequency in the three-phase coordinates shifts to $f_0 \pm f_d$. This explains the phenomenon in which an oscillation of 150 Hz in the *d*-*q* coordinates transforms into 100 Hz and 200 Hz in the three-phase coordinates.

### C. Added Values of This Paper

Based on the practical report, the following key conclusions are drawn to guide further theoretical investigations:

1) The oscillation is caused by dynamic interactions among functional outer control loops related to reactive power, rather than steady-state harmonics of converter switching. The steady-state harmonics caused by switching of three-level converters typically exhibit odd-numbered harmonic frequencies in three-phase coordinates, such as the 350 Hz component corresponded to the 7$^{\text{th}}$ harmonic. Therefore, the oscillations at 100 Hz and 200 Hz are dynamic oscillations and different from steady-state harmonics of the three-level converter [31].

2) Examinations of a single ESS have been conducted on both a hardware-in-the-loop (HIL) platform and a practical power system, demonstrating that a single ESS remains stable under all conditions. However, instability occurs when multiple ESSs are integrated, indicating that the instability is caused due to dynamic interactions among large-scale ESSs. Therefore, the stability analysis in this paper will be conducted from a system-level perspective instead of getting involved in an ESS.

3) The PLL of ESSs exhibits well-damped dynamics under strong grid connections, and the active power remains close to zero during the oscillation. Therefore, the PLL and active power outer control loop are not the causes. Engineers at the Electric Power Research Institute analyze the impedance characteristics of an ESS based on the HIL examination platform. The findings confirm that the functional outer control loop for reactive power can generate a damped oscillation at 150 Hz and change the impedance characteristics of an ESS around 150 Hz. Thus, it is identified as the source of oscillation frequency. This indicates that the oscillation is a novel phenomenon occurring under strong grid conditions and related to reactive power.

Therefore, this study aims to address the following issues:

1) How large-scale ESSs interact with each other, and why dynamic interactions among ESSs result in system instability.

2) How to plan the scale of ESSs by considering stability limitations of the whole power system.

3) Whether the conclusions obtained from this specific case (parallel connection) are applicable to other types of networks.

## III. Transfer Function of Grid-connected ESSs under Confidentiality Constraints

### A. Stability Analysis Scheme under Confidentiality Constraints

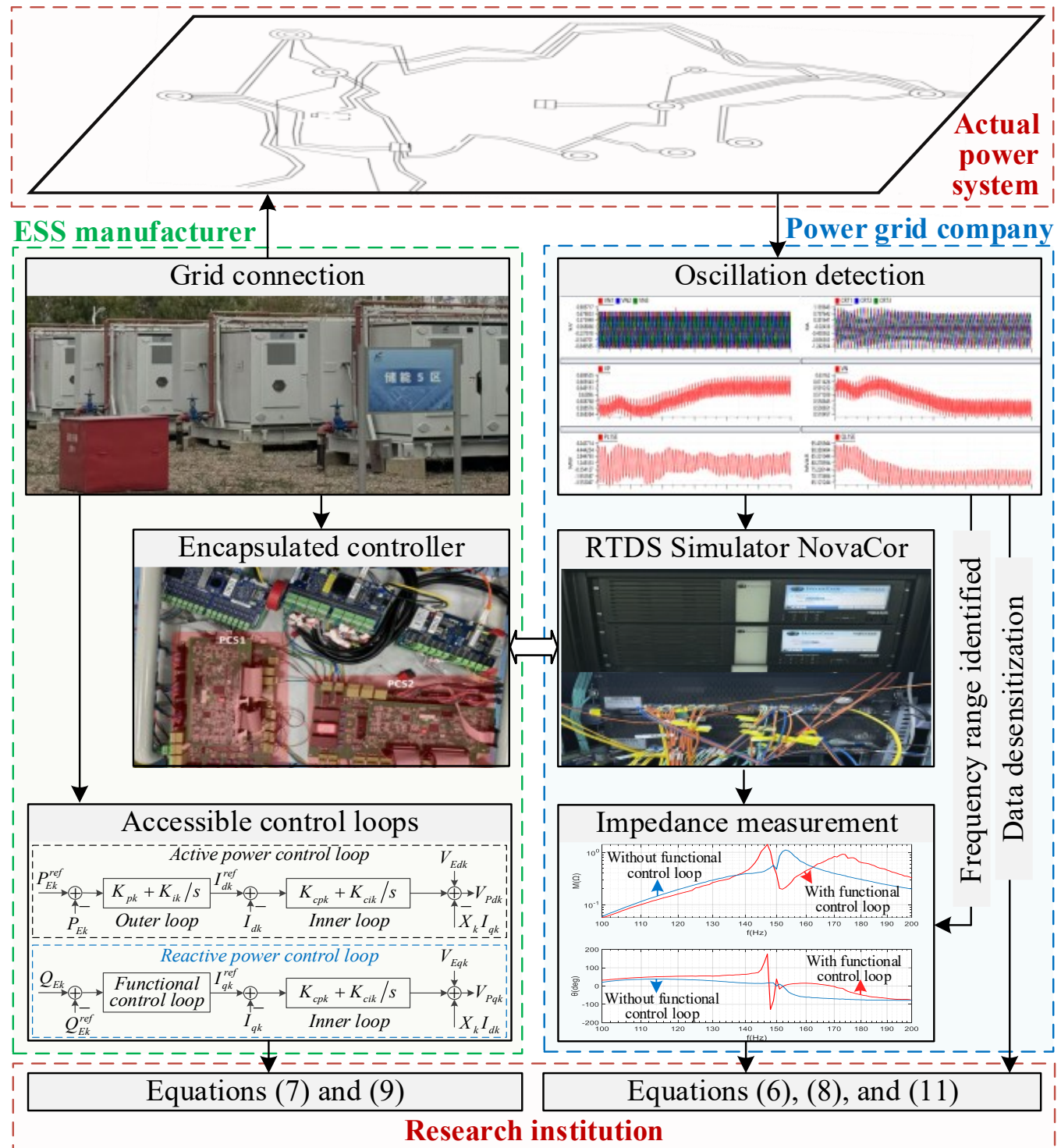

Fig. 8. Illustration of stability analysis scheme under confidentiality constraints.

The stability analysis of the oscillation involves three key stakeholders: the ESS manufacturer, the power grid company, and the research institution. The ESS manufacturer cannot fully disclose control configurations and parameters to the power grid company and the research institution, as such information is core intellectual property. Meanwhile, the power grid company cannot provide complete data outside the oscillation frequency range and power grid regions, as such information is irrelevant to such oscillation and may threaten power grid security. The research institution must analyze and address the oscillation from the perspective of the power system instead of the device, while adhering to the confidentiality constraints imposed by the ESS manufacturer and the power grid company. To achieve this target, a new scheme based on HIL for stability analysis is proposed and illustrated in Fig. 8.

In Fig. 8, the controller of the ESSs is encapsulated in such a way that its dynamic characteristics are preserved, while the internal information remains inaccessible. Next, the power grid company will measure the impedance curve of the controller using an HIL platform. If the impedance curve fails to meet the requirements for grid connection, meaning that the oscillation originates from the controller itself, the ESS manufacturer must revise the controller until it passes the test. Otherwise, by combining the controller impedance curve with measurements from the grid, the primary oscillation frequency range and the affected grid regions will be identified. These datasets will then be desensitized to reflect only the external dynamic behavior and not internal details. They include the equivalent transfer function of the ESS controller derived from the impedance curve and the equivalent transmission line impedance. Based on these, the research institution will conduct oscillation analysis and propose practical mitigation methods from the system-level perspective.

*B. Transfer Function of ESSs at Oscillation Frequency*

By fitting the difference between the impedance curves with and without the reactive power functional outer control loop, the equivalent transfer function of the functional outer control loop of the $k^{th}$ ESS within the identified frequency range (around 150 Hz, 3 p.u.) can be obtained, such method was also adopted in [32]. In this paper, the focused dynamics are represented by a second-order transfer function, which was conducted by the Electric Power Research Institute of State Grid Jiangsu Electric Power Co., Ltd., with contributions from Zijun Bin and Peng Li. The obtained expression for the equivalent transfer function is

$$\Delta I_{qk}^{ref} = \kappa_k(s)\Delta Q_{Ek} = \frac{K_{qk}s}{s^2+\zeta_q s+\omega_q{}^2}\Delta Q_{Ek} \tag{6}$$

where, $\kappa_k(s)$ is the transfer function of functional outer control loop of the $k^{th}$ ESS. In $\kappa_k(s)$, $K_{qk}$ is an amplification coefficient, $\zeta_q$ is a damping coefficient as designed by engineers, and $\omega_q$ is the concerned frequency, i.e., 3 p.u. Δ represents the variation in the variables.

It is important to note that $K_{qk}$ is not constant but is related to the amplitude of the reactive power. As explained by engineers, this design allows the functional outer control loop to be more sensitive under conditions of heavy reactive power support. Mathematically, it can be expressed as $K_{qk} = K_{q0}*|Q_{Ek}|$, $K_{q0}$ is constant.

The transfer function of the inner current control loop can be obtained from [33] as

$$\Delta I_{dk} = G_{ik}(s)\Delta I_{dk}^{ref},\ \Delta I_{qk} = G_{ik}(s)\Delta I_{qk}^{ref} \tag{7}$$

where $G_{ik}(s) = \frac{\omega_0\left(K_{cpk}s+K_{cik}\right)}{X_k s^2+\omega_0\left(K_{cpk}s+K_{cik}\right)}$, $\omega_0$ is the fundamental frequency as 1 p.u., which is 1 p.u. in per-unit terms or 314.16 rad/s in nominal value.

The PLL dynamics are considered to be ideal, as the grid connection is strong, meaning that the PLL instantly tracks the direction of $V_{Ek}$. The direction of $V_{Ek}$ lies along the $d$-axis of the ESS $d$-$q$ coordinates (i.e., $V_{Eq} = 0$, $V_{Ed} = V_E$) [34], [35]. The transfer function of the reactive power control loop can be derived by substituting (6) and (7) into the expression $\Delta Q_{Ek} = -I_{qk}\Delta V_{Edk} - V_{Edk}\Delta I_{qk}$, yielding:

$$\Delta I_{qk} = H_{qk}(s)\Delta V_{Edk} \tag{8}$$

where $H_{qk}(s) = -\frac{I_{qk}G_{ik}(s)\kappa_k(s)}{1+V_{Edk}G_{ik}(s)\kappa_k(s)}$.

Similarly, the transfer function of the active power control loop is obtained as (detailed derivations can be found in [33])

$$\Delta I_{dk} = H_{dk}(s)\Delta V_{Edk} \tag{9}$$

where $H_{dk}(s) = -\frac{I_{dk}G_{ik}(s)\left(K_{pk}s+K_{ik}\right)}{s+V_{dk}G_{ik}(s)\left(K_{pk}s+K_{ik}\right)}$.

Thus, from (8) and (9), the transfer function of the $k^{th}$ ESS in local $d$-$q$ coordinates can be obtained. However, the external power system is established in global $x$-$y$ coordinates. Using the transformation between $d$-$q$ and $x$-$y$ coordinates [36], the ESS transfer function in $x$-$y$ coordinates can be obtained as

$$\begin{bmatrix}\Delta I_{xk}\\ \Delta I_{yk}\end{bmatrix} = \mathbf{E}_k(s)\begin{bmatrix}\Delta V_{Exk}\\ \Delta V_{Eyk}\end{bmatrix},\ k\in[1,M] \tag{10}$$

where

$$\mathbf{E}_k(s) = \mathbf{T}\begin{bmatrix}H_{dk}(s)0\\ H_{qk}(s)0\end{bmatrix}\left(\mathbf{T}^{-1}+\dot{\mathbf{T}}^{-1}\begin{bmatrix}V_{Exk} & 0\\ 0 & V_{Eyk}\end{bmatrix}\mathbf{T}_\theta\right)+\dot{\mathbf{T}}\begin{bmatrix}I_{dk} & 0\\ 0 & I_{qk}\end{bmatrix}\mathbf{T}_\theta,$$

$$\mathbf{T} = \begin{bmatrix}\cos\theta_k & -\sin\theta_k\\ \sin\theta_k & \cos\theta_k\end{bmatrix},\quad \mathbf{T}^{-1} = \begin{bmatrix}\cos\theta_k & \sin\theta_k\\ -\sin\theta_k & \cos\theta_k\end{bmatrix},\quad \dot{\mathbf{T}}^{-1} = \begin{bmatrix}-\sin\theta_k & \cos\theta_k\\ -\cos\theta_k & -\sin\theta_k\end{bmatrix}$$

$$\dot{\mathbf{T}} = \begin{bmatrix}-\sin\theta_k & -\cos\theta_k\\ \cos\theta_k & -\sin\theta_k\end{bmatrix},\quad \mathbf{T}_\theta = \frac{1}{V_{Ek,0}^2}\begin{bmatrix}-V_{Eyk} & V_{Exk}\\ -V_{Eyk} & V_{Exk}\end{bmatrix}.$$

For the practical power system, M ESSs are interconnected in parallel and connected to the external power system through an inductance $L_T$. The impedance matrix of AC transmission line can be represented in $x$-$y$ coordinates as

$$\Delta\mathbf{V}_{xy} = \left(sL_T\times\mathbf{I}_{2\times2}\right)\Delta\mathbf{I}_{xy} \tag{11}$$

where $\Delta\mathbf{I}_{xy} = [\Delta I_{x1}, \Delta I_{y1}]^T = \ldots = [\Delta I_{xM}, \Delta I_{yM}]^T$, $\Delta\mathbf{V}_{xy} = [\Delta V_{x1}, \Delta V_{y1}]^T = \ldots = [\Delta V_{xM}, \Delta V_{yM}]^T$. **I** is an identity matrix.

Combining (10) and (11), the transfer function of the practical power system with parallel-connected ESSs is

$$sML_T\mathbf{E}_k(s) = \mathbf{I}_{2\times2} \tag{12}$$

## IV. Instability Mechanism of the Practical Oscillation: Damping Analysis and Improvement

*A. Superimposed Effects of Practical Parallel-connected ESSs on the Oscillation Damping*

The stability of the power system can be determined by the solutions of (12), particularly the dominant oscillation mode. Given that the observed oscillation is primarily caused by dynamic interactions among the reactive power functional outer control loops of multiple ESSs, and considering that the active power is close to zero and can therefore be ignored, the characteristic equation of the power system connected with M parallel-connected ESSs can be derived from (12) as

$$s^2 + \zeta_q s + \omega_q^{\ 2} + d_\Sigma(s) = 0 \tag{13}$$

where $s^2 + \zeta_q s + \omega_q^2$ represents the second-order oscillation loop dominated by the reactive power functional outer control loop, as defined in (11). $d_\Sigma(s)$ denotes the transfer function of the other dynamics, including the dynamics from other control loops and the dynamic interactions among ESSs. The voltage angle is set as zero for simplification.

The expression of $d_\Sigma(s)$ is given by

$$\begin{aligned} d_\Sigma(s) &= d_N(s) + d_S(s) \\ &= \underbrace{V_{Ek} G_{ik}(s) K_{qk} s}_{\text{Normal effect}} \underbrace{- M^2 V_{Ek}^{\ -1} (L_T I_{qk})^2 G_{ik}(s) K_{qk} s^3}_{\text{Superimposed effect}} \end{aligned} \tag{14}$$

The normal effect from $d_N(s)$ is only affected by the control parameters and operational states of ESSs, irrespective of the number of ESSs. It can be assessed once the parameters are identified. By contrast, the superimposed effect from $d_S(s)$ represents the impact that intensifies as the number of ESSs increases. Unlike the normal effect, the superimposed effect varies even when the control parameters and operational states remain constant. This becomes more pronounced as the scale of ESSs expands, necessitating particular attention and thus becoming the primary focus of this study.

It can be observed that the dynamics of functional outer control loop are irrelated to the external power system if $d_S(s) = 0$. This corresponds to an ideal scenario in which the ESS is connected to an infinite bus. In this case, the dynamics of the functional outer control loop are the same as those of the designed control dynamics. However, if grid-side dynamics are considered, the dynamics of functional outer control loop are affected by the dynamic interactions among ESSs, i.e., $d_S(s)$, since $d_S(s) \neq 0$. It reflects the significance of the superimposed effect of ESSs.

In addition to the number of ESSs, the superimposed effect is influenced by several factors, including external parameters such as the inductance of transmission line and the amplitude of reactive power, as well as internal parameters such as the proportional coefficients $K_{qk}$. To numerically assess the impact of the normal and superimposed effects on system stability, the analysis primarily focuses on the oscillation damping, and the stability criterion can be expressed as:

$$\zeta_q + \zeta_\Sigma = \zeta_q + \zeta_N + \zeta_S > 0 \tag{15}$$

where $\zeta_N$ and $\zeta_S$ represent the damping provided by the normal and superimposed effects of ESSs, respectively.

*B. Damping Analysis for Quantifying the Superimposed Effects of ESSs*

The virtual damping method [37] is adopted to numerically evaluate the damping $\zeta_S$ provided by the superimposed effect of ESSs and, consequently, to determine system stability using (15). Here, the critical stability condition is considered, and thus $s = j\omega_q$. Therefore, (13) can be expressed as:

$$s^2 + \zeta_q s + \omega_q^{\ 2} + d_\Sigma(j\omega_q) = 0 \tag{16}$$

where $d_\Sigma(j\omega_q)$ is a complex value, that is, $\mathrm{Re}[d_\Sigma(j\omega_q)] + j\mathrm{Im}[d_\Sigma(j\omega_q)]$. Re[ ] and Im[ ] denote the real and imaginary parts of the complex value, respectively.

Subsequently, $d_\Sigma(j\omega_q)$ can further be converted to a value in the complex frequency domain with $s = j\omega_q$ as:

$$d_\Sigma(j\omega_q) = \zeta_\Sigma s + C = (\zeta_N + \zeta_S)s + C \tag{17}$$

To make the expression of $\zeta_N$ and $\zeta_S$ more observable, it is simplified by considering $V_{Ek} \approx 1$ p.u., yielding

$$\begin{aligned} \zeta_N &= \frac{\mathrm{Im}\left[d_N(j\omega_q)\right]}{\omega_q} = -\frac{K_{qk}}{X_k \omega_q^{\ 2}} \\ \zeta_S &= \frac{\mathrm{Im}\left[d_S(j\omega_q)\right]}{\omega_q} = -K_{qk}\frac{(ML_T I_{qk}\omega_q)^2}{X_k \omega_q^{\ 2}} \end{aligned} \tag{18}$$

where the high-order terms in $G_{ik}(s)$ dominate, and thus the low-order terms are neglected since $\omega_q^2 >> 1$.

Using (18), the impact of superimposed effect of ESSs on the oscillation damping can be numerically evaluated through the index $\zeta_S$. A higher $\zeta_S$ indicates stronger dynamic interactions and a superimposed effect, resulting in more pronounced effects on system stability. In practical power systems in which ESSs are connected in parallel, the strength of the superimposed effect is positively correlated with the number of ESSs connected in parallel. As the number of ESSs increases, the strength of the superimposed effect also increases.

Furthermore, the control parameter $K_{qk}$ also plays a critical role in determining the strength of the superimposed effect. When $K_{qk} = 0$, the ESSs do not engage in the external power system dynamics, implying that the strength of superimposed effect is negligible. As $K_{qk}$ increases, the superimposed effect becomes more significant, potentially leading to substantial changes in the oscillation damping. This highlights a critical trade-off: while the functional outer control loop facilitates the adjustment of power system operating states, it simultaneously introduces additional dynamic interactions that may pose potential risks to system stability. Therefore, the careful optimization of $K_{qk}$ is essential to balance the benefits of ESS integration with the mitigation of adverse dynamic effects.

*C. Identification of Major Factors Impacting the Damping*

The oscillation damping at the oscillation frequency ($\omega_q = 3$ p.u.) can be calculated from (13) and (18) as

$$\varepsilon_q = \frac{\zeta_q + \zeta_\Sigma}{2} = \frac{\zeta_q}{2} - K_{qk}\frac{1 + (ML_T I_{qk}\omega_q)^2}{2X_k \omega_q^{\ 2}} \tag{19}$$

*1) Reactive Power Amplitude and Direction.*

It can be seen from (19) that the oscillation damping varies with changes in the reactive power amplitude (corresponding to $I_{qk}$). As inductive reactive power increases, $I_{qk}$ increases in the positive direction, causing an increased amplitude of $I_{qk}^2$, which reduces the oscillation damping. Similarly, $I_{qk}^2$ increases and the oscillation damping decreases when capacitive reactive power increases. Therefore, the impact of reactive power output from ESSs on the oscillation damping mainly depends on its amplitude, regardless of its direction.

*2) Integration Number of ESSs.*

The derivative of the oscillation damping with respect to the number of ESSs can be obtained from (19) as

$$\frac{\partial \varepsilon_q}{\partial M} = -2MK_{qk}\frac{(L_T I_{qk}\omega_q)^2}{2X_k\omega_q^{\ 2}} \tag{20}$$

As the number of ESSs increases, the impact of dynamic interactions among ESSs on the system stability is amplified, reducing the oscillation damping. It can be observed in (20) that the impact of increasing the number of ESSs on stability is non-linear: the more ESSs are connected, the more significant the effects on oscillation damping become. This clarifies why the superimposed effect of ESSs has a significant influence on system stability, particularly when the ESS scale is large. It also suggests that the stability margin may degrade rapidly as the ESS scale expands. Consequently, the maximum number of ESSs must be carefully constrained once the rated power of the ESSs has been established, to ensure system stability and mitigate potential risks associated with dynamic interactions among ESSs.

*3) Reactive Power Control Loop Parameters.*

Two control parameters, $K_{qk}$ and $\zeta_q$, affect the oscillation damping. The self-stability of ESSs mainly depends on $\zeta_q$ and $\zeta_N$, meaning that a larger $\zeta_q$ allows ESSs to better defend against negative effects from the dynamic interactions. From (19), the relationship between oscillation damping and $\zeta_q$ is linear in nature.

$$\frac{\partial \varepsilon_q}{\partial K_{qk}} = -\frac{1+(ML_T I_{qk}\omega_q)^2}{2X_k\omega_q^{\ 2}} \tag{21}$$

In (21), the impact of $K_{qk}$ on the oscillation damping depends on external conditions, including the ESS number, reactive power, and inductance of the transmission line. Increasing $K_{qk}$ is always negative to the oscillation damping. These effects are amplified when the ESS scale is large. Therefore, the value of $K_{qk}$ should be kept small to reduce the effects from external factors on the oscillation damping.

*4) Inductance of Transmission Line*

As can be observed from (19) and (20), both the transmission line inductance and the ESS number have similar effects on system dynamics. Specifically, the oscillation damping decreases as the transmission line inductance increases. However, a decrease in the ESS number tends to counteract the effect of larger transmission line inductance. In other words, the system is more prone to instability with a higher transmission line inductance than with a lower inductance.

*D. Solutions for Improving the Oscillation Damping*

*1) Constraint on the Scale of ESSs*

To ensure that $\varepsilon_q > 0$, the constraints on the maximum number and reactive power amplitude of ESSs are derived from (19) as follows:

$$\begin{gathered}\frac{\zeta_q}{2} - K_{qk}\frac{1+(ML_T I_{qk}\omega_q)^2}{2X_k\omega_q^{\ 2}} > 0 \\ \Downarrow \\ I_{qk}^{\ 2} < \frac{\zeta_q X_k\omega_q^{\ 2} - K_{qk}}{K_{qk}(ML_T\omega_q)^2} \\ \Downarrow\ Q_{Ek} = I_{qk}V_{Ek} \\ M\left|Q_{Ek}\right| < \frac{V_{Ek}}{L_T\omega_q}\sqrt{\frac{\zeta_q X_k\omega_q^{\ 2} - K_{qk}}{K_{qk}}}\end{gathered} \tag{22}$$

From (22), reducing either the number of ESSs or the rated reactive power of each ESS can improve stability, which should be carefully considered during the planning stage.

*2) Increase the SCR of Power System*

From the perspective of the power grid company, it is their duty to provide sufficiently strong grid connections. Although the SCR in this practical case was relatively high, the scale of ESSs suggests that the grid connection strength needs to be further improved. The minimum SCR for the grid connection should satisfy the following constraint as

$$SCR > MI_{qk}\omega_q\sqrt{\frac{K_{qk}}{\zeta_q X_k\omega_q^{\ 2} - K_{qk}}} \tag{23}$$

*3) Additional Damping Control Loop.*

The above solutions are recommended primarily for the planning stage rather than real-time operation. Moreover, the control parameters are confidential and inaccessible to external engineers, and thus cannot be adjusted in practice. Therefore, an additional control loop added to the reactive power control loop is proposed to improve the oscillation damping, which is

$$\Delta I_{qk}^{ref} = \kappa_k(s)\left(\Delta Q_{Ek} - K_{dm}\Delta I_{qk}^{ref}\right) \tag{24}$$

where $K_{dm}$ is a damping coefficient of the additional damping control loop.

Using this additional control, the improved damping is

$$\varepsilon_q = K_{qk}\left(\frac{\zeta_q K_{dm}}{2} - \frac{1+(ML_T I_{qk}\omega_q)^2}{2X_k\omega_q^{\ 2}}\right) \tag{25}$$

With the proposed method, the oscillation damping can be improved by increasing $K_{dm}$, effectively addressing instability issues caused by various factors and simplifying parameter adjustments for operational engineers.

## V. Generalization and Error Analysis

*A. Damping Analysis under Generalized Network*

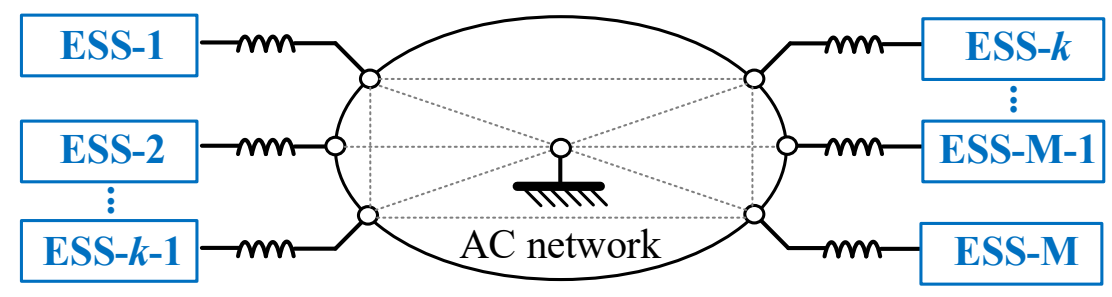

Fig. 9. Configuration of large-scale ESSs interconnected through a generalized network.

Consider a generalized scenario where M ESSs are connected to an external power system through a generalized network, as illustrated in Fig. 9. The admittance matrix of the AC network can be represented in the *x*-*y* coordinate system as

$$\Delta \mathbf{I}_{xy} = (\mathbf{Y}_{xy}(s) \otimes \mathbf{I}_{2\times 2})\Delta \mathbf{V}_{xy} \tag{26}$$

where $\mathbf{I}_{xy} = [I_{x1}, I_{y1}, \ldots, I_{xM}, I_{yM}]^T$, $\mathbf{V}_{xy} = [V_{x1}, V_{y1}, \ldots, V_{xM}, V_{yM}]^T$. $\mathbf{Y}_{xy}(s) = (sL_0\mathbf{L})^{-1}$. Here, $L_0$ represents the inductance per unit length of the transmission line, and $\mathbf{L}$ is a matrix composed of the lengths of the lines. The symbol $\otimes$ denotes the Kronecker product of two matrices.

From (10) and (26), the transfer function of the power system connected with M ESSs is given by

$$Diag(\mathbf{E}_k(s)) = \mathbf{Y}_{xy}(s) \otimes \mathbf{I}_{2\times 2} \tag{27}$$

where *Diag* indicates a diagonal matrix.

Multiplying $\mathbf{W}$ and $\mathbf{W}^{-1}$ to (27), yields

$$\begin{gathered}\left(\mathbf{W}\mathbf{I}_{M\times M}\mathbf{W}^{-1}\right) \otimes \mathbf{E}_k(s) = (\mathbf{W}\mathbf{Y}_{xy}(s)\mathbf{W}^{-1}) \otimes \mathbf{I}_{2\times 2} \\ \Downarrow \\ Diag(\mathbf{E}_k(s)) = Diag(\rho_k(s)) \otimes \mathbf{I}_{2\times 2}\end{gathered} \tag{28}$$

where, $\mathbf{W}$ and $\mathbf{W}^{-1}$ are matrixes that are composed of left and right eigenvectors of $\mathbf{Y}_{xy}(s)$, respectively. $\rho_k(s)$ represents a transfer function corresponding to the $k^{th}$ eigenvalue of $\mathbf{Y}_{xy}(s)$, such that $Diag(\rho_k(j\omega_q)) = \mathbf{W}\mathbf{Y}_{xy}(j\omega_q)\mathbf{W}^{-1}$.

Therefore, all oscillation modes can be obtained by solving M equations in (28). This differs from the scenario where all ESSs are parallel-connected, in which the dominant oscillation mode is an M-repeated root. In this generalized case, the M oscillation modes overlap considering that $\rho_k(s)$ are generally different. Therefore, the system stability is mainly determined by the dominant oscillation mode. If the dominant oscillation mode corresponds to $\rho_M(s)$, it can be calculated by

$$\mathbf{E}_M(s) = \rho_M(s) \times \mathbf{I}_{2\times 2} \tag{29}$$

Thus, the characteristic equation corresponding to dominant oscillation mode of the generalized power system can be derived from (13) and (29) as

$$s^2 + \zeta_q s + \omega_q^2 + d_\Sigma(s) = 0 \tag{30}$$

where $d_\Sigma(s) = V_{Ek}G_{ik}(s)K_{qk}s - V_{Ek}^{-1}\left(\rho_M(s)^{-1}I_{qk}\right)^2 G_{ik}(s)K_{qk}s$.

The impact of superimposed effect of ESSs on the oscillation damping can be quantified using the virtual damping method, expressed as

$$\zeta_\Sigma = \frac{\mathrm{Im}\left[d_\Sigma(j\omega_q)\right]}{\omega_q} = -K_{qk}\frac{1 + (\rho_M{}^{-1}I_{qk}\omega_q)^2}{X_k\omega_q{}^2} \tag{31}$$

where $\rho_M$ is the coefficient of $\rho_M(s)$.

Considering the similarity between (18) and (31), it is evident that the damping analysis results drawn for parallel-connected ESSs are also applicable to generalized power systems, which are not repeated here.

*B. Generalization of the Superimposed Effect of ESSs*

Furthermore, the correctness of the superimposed effect of ESSs in generalized scenarios is verified as follows.

When an (M+1)$^{th}$ ESS is added to the original power system, the admittance matrix of the network is affected, resulting in an increase in the dimension of $(\mathbf{Y}_{xy}(s) \otimes \mathbf{I}_{2\times 2})$ by two orders and an increase in the order of $\mathbf{Y}_{xy}(s)$ by one. For clarity, the original and modified admittance matrices are denoted as $\mathbf{Y}_O$ and $\mathbf{Y}_N$, respectively, yielding

$$\mathbf{Y}_O = \frac{1}{sL_0}\begin{bmatrix}\mathbf{L}^{-1} & 0 \\ 0 & 0\end{bmatrix},\ \mathbf{Y}_N = \frac{1}{sL_0}\begin{bmatrix}\mathbf{L}^{-1} + \mathbf{l}_{11} & \mathbf{l}_{12} \\ \mathbf{l}_{21} & l_{22}\end{bmatrix} \tag{32}$$

Based on (32), it can have

$$\rho_{MO} = \frac{1}{sL_0}\begin{bmatrix}\mathbf{w}_1 & w_2\end{bmatrix}\begin{bmatrix}\mathbf{L}^{-1} & 0 \\ 0 & 0\end{bmatrix}\begin{bmatrix}\mathbf{w}_1 & w_2\end{bmatrix}^T \tag{33}$$

where $\rho_{MO}$ denotes the eigenvalue of $\mathbf{Y}_O$ corresponding to the dominant oscillation mode before connecting the (M+1)$^{th}$ ESS. $[\mathbf{w}_1\ w_2]$ is a characteristic vector corresponding to $\rho_{MO}$.

From (32) and (33), the eigenvalue of $\mathbf{Y}_N$ corresponding to the dominant oscillation mode after connecting the (M+1)$^{th}$ ESS can be approximately calculated as

$$\rho_{MN} = \frac{1}{sL_0}\begin{bmatrix}\mathbf{w}_1 & w_2\end{bmatrix}\begin{bmatrix}\mathbf{L}^{-1} + \mathbf{l}_{11} & \mathbf{l}_{12} \\ \mathbf{l}_{21} & l_{22}\end{bmatrix}\begin{bmatrix}\mathbf{w}_1 & w_2\end{bmatrix}^T \tag{34}$$

Comparing (34) with (33), yields

$$\begin{aligned}\rho_{MN} &= \frac{\mathbf{w}_1\mathbf{L}^{-1}\mathbf{w}_1{}^T + \mathbf{w}_1\mathbf{l}_{11}\mathbf{w}_1{}^T + w_2\mathbf{l}_{21}\mathbf{w}_1{}^T + \mathbf{w}_1\mathbf{l}_{12}w_2 + w_2{}^2 l_{22}}{sL_0} \\ &= \rho_{MO} + \frac{\mathbf{w}_1\mathbf{l}_{11}\mathbf{w}_1{}^T + w_2\mathbf{l}_{21}\mathbf{w}_1{}^T + \mathbf{w}_1\mathbf{l}_{12}w_2 + w_2{}^2 l_{22}}{sL_0} > \rho_{MO}\end{aligned} \tag{35}$$

All variables involved in (35) are positive, which confirms that the eigenvalue of the new admittance matrix increases when an additional ESS is connected. This increase in $\rho_M(s)$ in (30) is analogous to the increase in M in (19), demonstrating that the superimposed effect of ESSs is generalized and not limited to specific connection configurations. Consequently, the number of ESSs must be carefully constrained in generalized networks to ensure system stability.

*C. Impact of ESS Model Errors on Oscillation Damping*

The ESS models may not be perfectly identical, and thus model errors may affect the result of oscillation damping. Let the proportional error of the $k^{th}$ ESS model be denoted as $e_k(s)$. Then, (27) changes to

$$\begin{gathered} Diag(e_k(s)) \otimes \mathbf{E}_k(s) = \mathbf{Y}_{xy}(s) \otimes \mathbf{I}_{2\times 2} \\ \Downarrow \\ \mathbf{I}_{M\times M} \otimes \mathbf{E}_k(s) = \left( Diag(e_k(s))^{-1} \mathbf{Y}_{xy}(s) \right) \otimes \mathbf{I}_{2\times 2} \end{gathered} \tag{36}$$

By comparing (36) with (28), we observe that the impact of ESS model errors is equivalent to proportionally modifying the network dynamics.

For the dominant oscillation mode, (29) changes to

$$\mathbf{E}_M(s) = e_M(s)^{-1} \rho_M(s) \times \mathbf{I}_{2\times 2} \tag{37}$$

where $e_M(s)^{-1} = \mathbf{w}_1 Diag(e_k(s))^{-1} \mathbf{w}_1^T$, $\mathbf{w}_1 \mathbf{w}_1^T = 1$.

From (37), for a given power system, the ESS model errors are constant, that is $e_k(j\omega_q)$ is constant. Therefore, the qualitative impacts analyzed in the previous subsections remain valid. As affected by the ESS model errors, the variation on oscillation damping can be evaluated by

$$\Delta\zeta_\Sigma = -\frac{K_{qk}\,\mathrm{Re}[e_M(j\omega_q)](\rho_M{}^{-1} I_{qk}\omega_q)^2}{X_k \omega_q{}^2} \Delta\,\mathrm{Re}[e_M(j\omega_q)] \tag{38}$$

The superimposed effect of large-scale ESSs also persists since $|e_k(j\omega_q)|$ is around 1. However, for ESSs with significantly different control strategies, where differences in models cannot be treated as modeling errors, the conclusions regarding the superimposed effect may not apply, as that the superimposed effect is only justified for large integration of similar ESSs.

### D. Impact of SG Model Errors on Oscillation Damping

In Fig. 9, the external bulk power system is represented by an infinite bus, which is reasonable because its capacity is significantly larger than that of the ESSs. To enhance accuracy, the dynamics of the external power system, particularly its inertia and damping, are considered by representing it as a SG by the following transfer function

$$\mathbf{G}_g(s) = \frac{r_g + s l_g}{2H_g s^2 + D_g s + K_g} \otimes \mathbf{I}_{2\times 2} \tag{39}$$

where $r_g$ and $l_g$ represent the equivalent impedance of the SG, and $H_g$ and $D_g$ denote the inertia and damping coefficients of the SG, respectively.

As an example, if the equivalent SG is connected at node M+1, then, analogous to (34), the variation of the eigenvalue of $\mathbf{Y}_N$ associated with the dominant oscillation mode becomes

$$\rho_{MN} = \frac{[\mathbf{w}_1 \;\; w_2]}{sL_0} \begin{bmatrix} \mathbf{L}^{-1} + \mathbf{l}_{11} & \mathbf{l}_{12} \\ \mathbf{l}_{21} & l_{22} - \dfrac{(r_g + s l_g) s L_0}{2H_g s^2 + D_g s + K_g} \end{bmatrix} \begin{bmatrix} \mathbf{w}_1 \\ w_2 \end{bmatrix} \tag{40}$$

In the high-frequency range, the high-order items are major, and thus the error can be approximated as $l_g L_0/(2H_g)$. As affected by the SG model errors, the variation on oscillation damping can be evaluated by

$$\Delta\zeta_\Sigma = -\frac{2K_{qk}(I_{qk}\omega_q)^2}{X_k \omega_q{}^2 \rho_M{}^3} \left( w_2{}^2 \frac{l_g L_0}{2H_g} \right) \tag{41}$$

This implies that larger bulk-system inertia reduces the error. Considering that the Jiangsu power grid is one of the largest provincial grids in China, the resulting error is negligible.

## VI. Case Studies

### A. Model Verification

A power system connected with 8 ESSs is established in the SIMULINK, utilizing the official three-level VSC model [38] for the electromagnetic transient (EMT) simulation. The system parameters are the same as those listed in Table I, and the others are given in Table II.

Table II
Parameters of the study case

| Parameters | Values |
|---|---|
| Active power regulator gains | 0.1, 100 rad/s |
| Current regulator gains | 0.25, 100 rad/s |
| Proportion gain of functional outer control loop | 0.07 |
| Coefficients of functional outer control loop | 0.01 p.u. 3 p.u. |
| Transmission line impedance | 0.14 p.u. |
| Fundamental frequency | 50 Hz |
| Simulation step time | 2e-5 sec. |
| Control sample time | 1e-4 sec. |

The dominant oscillation mode is calculated from (12) to be 0.0022 + $j$3.0 p.u. (0.69 + $j$942.55 rad/s) when the inductive reactive power output from ESSs is 0.35 p.u., and to be 0.0033 + $j$3.0 p.u. (1.04 + $j$ 942.55 rad/s) when the capacitive reactive power output from ESSs is 0.35 p.u. These results indicate that the power system is unstable with a frequency of 150 Hz in the *d-q* and *x-y* coordinates when the reactive power amplitude reaches 0.35 p.u. However, when the reactive power amplitude decreases to 0.25 p.u., the power system regains stability. The oscillation modes are calculated as −0.40212 + $j$942.52 rad/s for inductive reactive power and −0.21067 + $j$942.52 rad/s for capacitive reactive power.

For verification, the EMT simulation results are illustrated in Figs. 10 and 11 for capacitive and inductive reactive power scenarios, respectively. Fast Fourier Transform (FFT) analysis results are provided in Fig. 12. In Fig. 10, the system operates stably at the initial condition of 0.2 s, where the reactive power is maintained at 0.25 p.u. However, at 0.2 s, when the reactive power is increased to 0.35 p.u., significant oscillation can be observed. The FFT analysis results, depicted in Fig. 12, reveal that the dominant oscillation frequency is 150 Hz in the *d-q* and *x-y* coordinates, while frequencies of 100 Hz and 200 Hz are observed in the three-phase coordinates. These results are consistent with the theoretical analysis, thereby confirming their accuracy. Similar conclusions can be drawn from Fig. 11 for the case of inductive reactive power, which is not reiterated here for brevity.

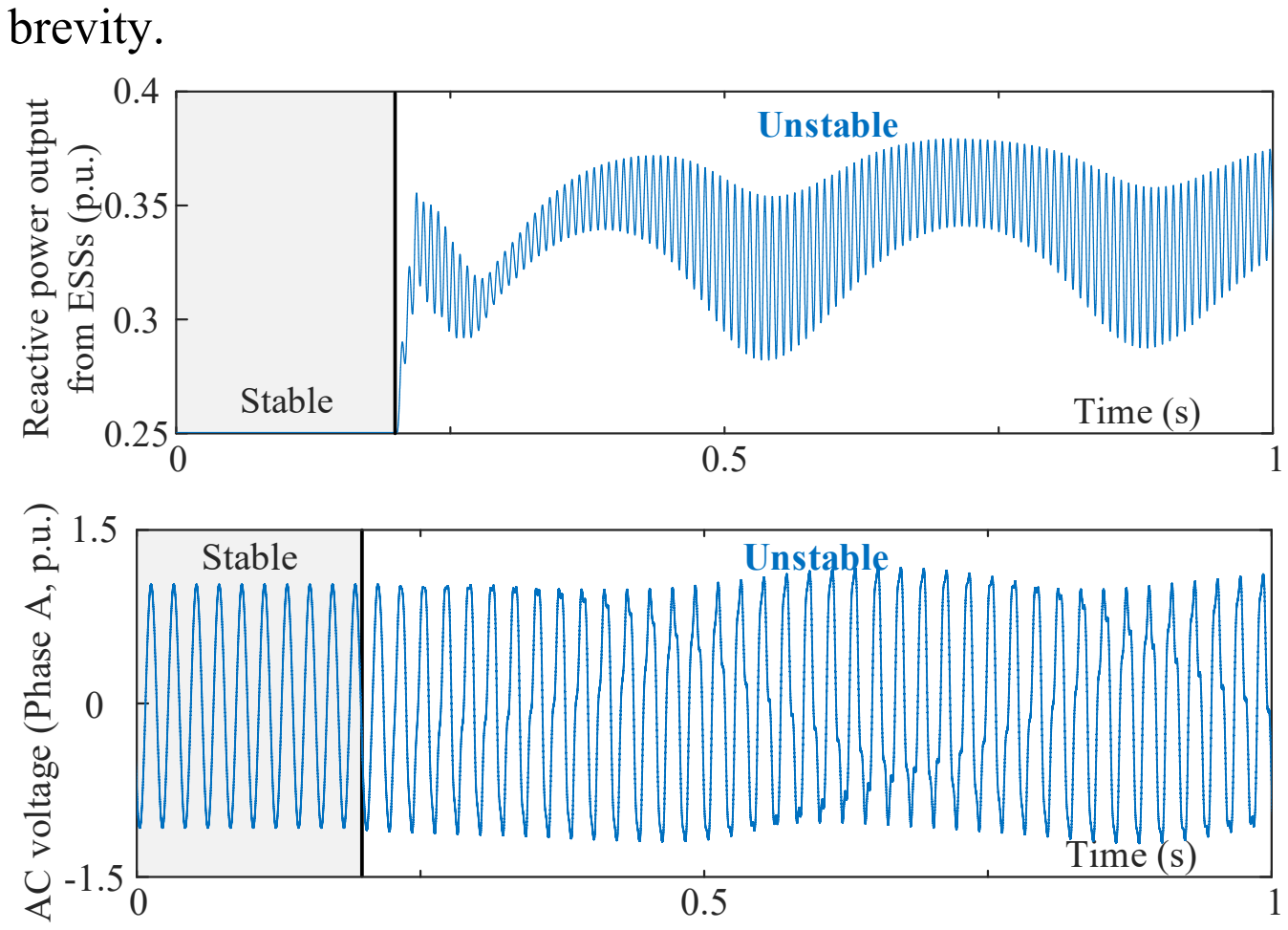

Fig. 10. Capacitive reactive power and AC voltage curves when reactive power output from ESSs varies from 0.25 p.u. to 0.35 p.u.

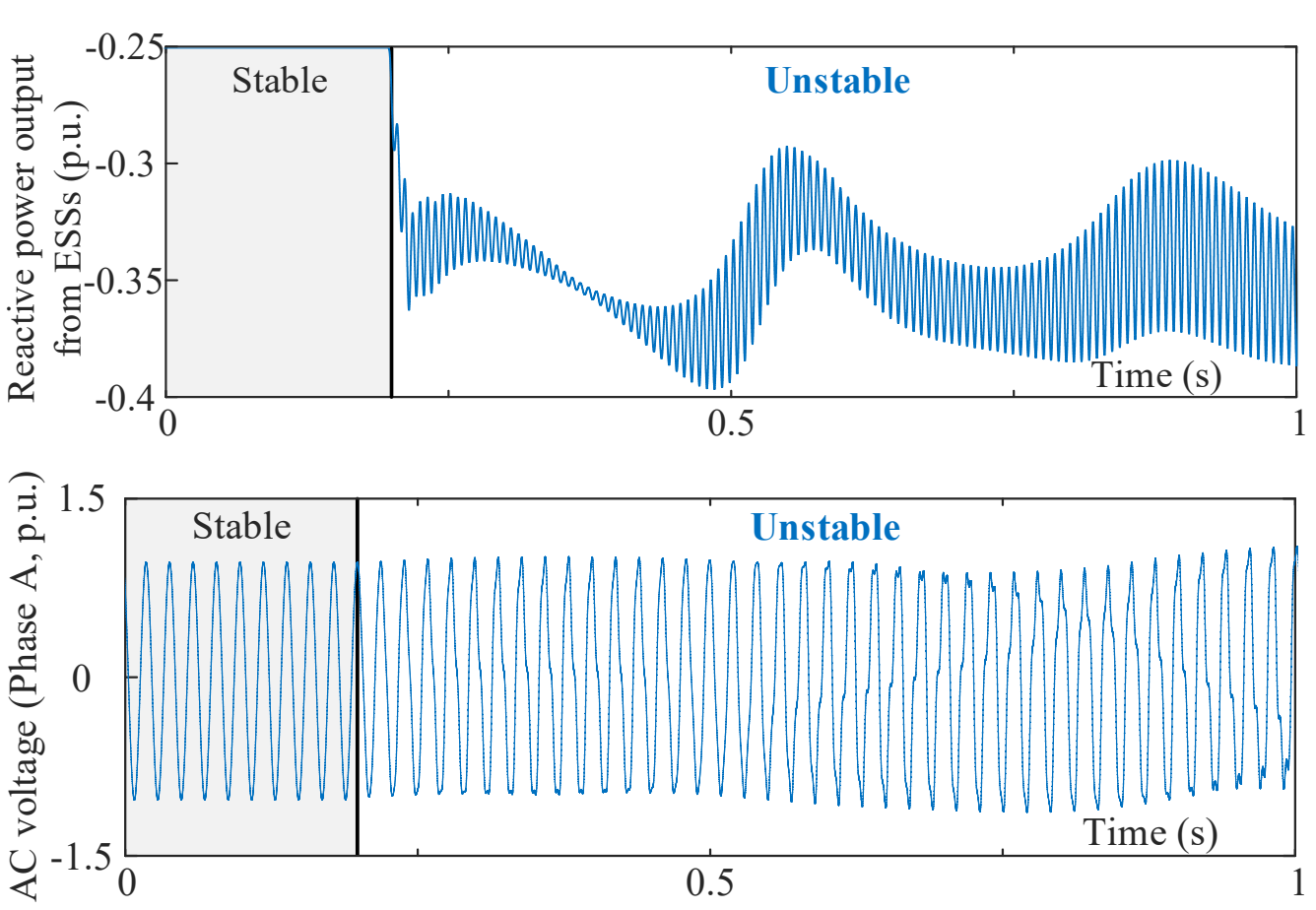

Fig. 11. Inductive reactive power and AC voltage curves when reactive power output from ESSs varies from 0.25 p.u. to 0.35 p.u.

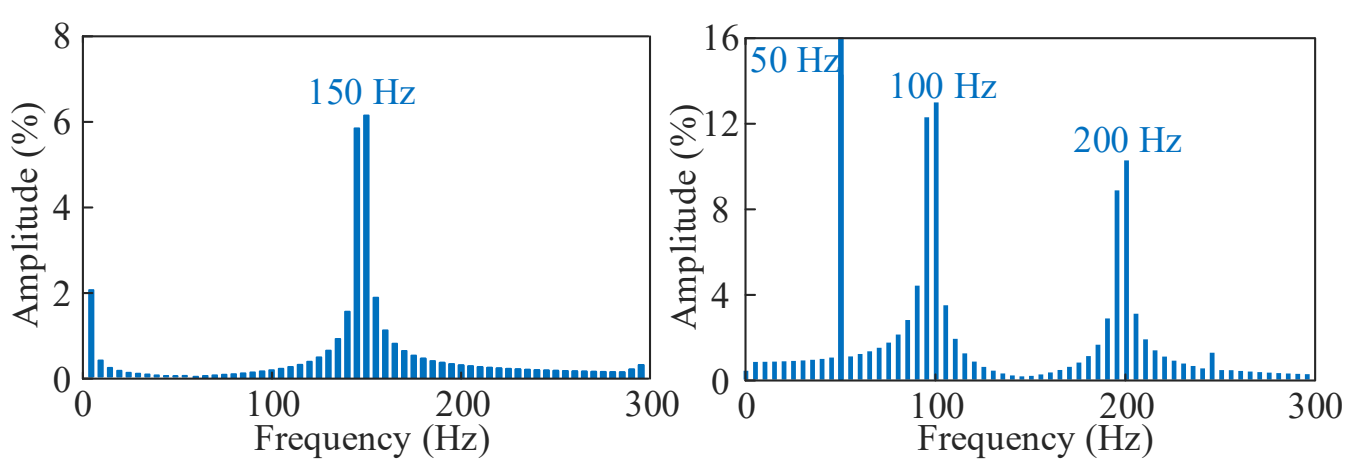

Fig. 12. FFT analysis results of the curves in Fig. 7: reactive power (left) and AC voltage (right).

These figures demonstrate that the power system remains stable when the reactive power amplitude is 0.25 p.u. However, as the amplitude increases to 0.35 p.u., the oscillation emerges. This indicates that the oscillation is closely associated with the reactive power control loop. Furthermore, the impact of reactive power output from ESSs on oscillation damping is primarily determined by its amplitude, irrespective of its direction (capacitive or inductive). These findings underscore the critical role of reactive power amplitude in influencing system stability and highlight the need for careful management of reactive power levels in ESS-integrated power systems.

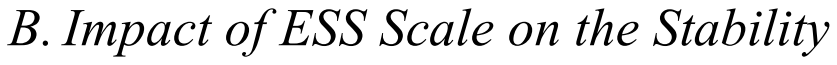

### *B. Impact of ESS Scale on the Stability*

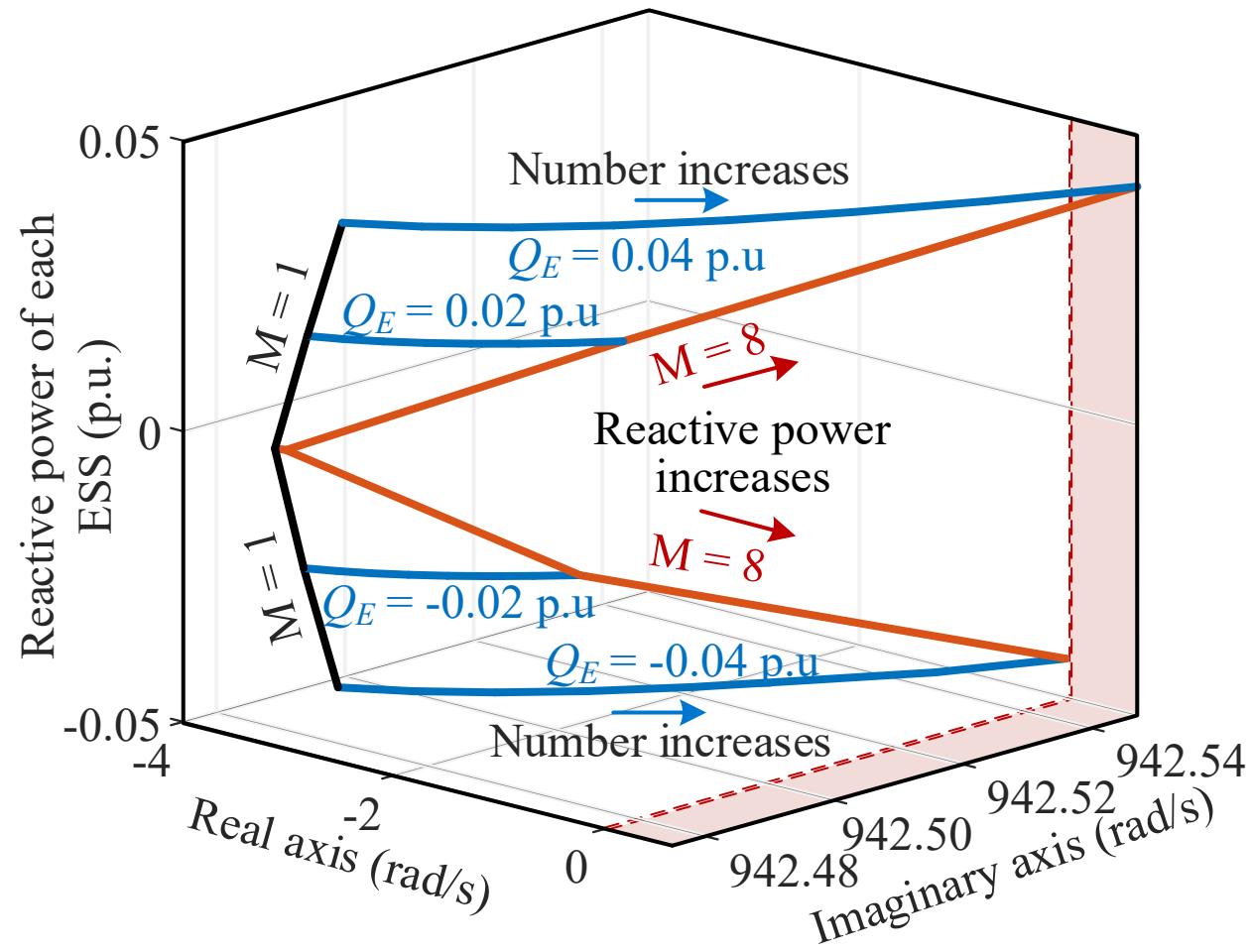

Fig. 13. Trajectories of the oscillation mode as the scale of ESSs increases.

The scale of ESSs is increased by either adding more ESSs or increasing the output power of each ESS. The oscillation mode is calculated as -3.13 + $j$942.48 rad/s when only one ESS is connected, and the output reactive power is close to 0 p.u. In this case, the power system is stable. However, as the reactive power amplitude increases, the oscillation damping decreases. As indicated by the trajectory highlighted in black in Fig. 13, the oscillation damping decreases from -3.13 rad/s to -2.69 rad/s and -2.70 rad/s as the reactive power varies from 0 p.u. to 0.04 p.u. and -0.04 p.u., respectively. Moreover, increasing the output power of each ESS causes more significant impacts on stability when more ESSs are connected. As indicated by the trajectory highlighted in red in Fig. 13, the oscillation mode moves to 0.55 + $j$942.54 rad/s and 0.49 + $j$942.54 rad/s as the reactive power of 8 ESSs increases to 0.32 p.u. and -0.32 p.u., respectively.

Similarly, increasing the ESS number also reduces stability, as shown by the blue trajectories in Fig. 13. For example, the oscillation damping decreases from -2.69 rad/s to 0.55 rad/s as the number of ESSs increases from 1 to 8, while the reactive power of each ESS remains 0.04 p.u. Likewise, the oscillation damping decreases from -2.70 rad/s to 0.49 rad/s when the ESS number increases from 1 to 8, with each ESS outputting -0.04 p.u. The results demonstrate that the oscillation damping decreases as the scale of ESSs improves, either through an increase in their number or output reactive power.

The EMT simulation results under different operational states are shown in Fig. 14, which demonstrates the relationship between oscillation damping and the reactive power output from the ESSs. It can be observed that the oscillation damping decreases as the reactive power of ESSs increases, regardless of whether they are operating in the inductive or capacitive mode. Notably, the power system loses stability when the total amplitude of output reactive power from all ESSs reaches 0.32 p.u.

To illustrate it more clearly, the results presented in Fig. 14 (a) are focused on. Initially, when the output reactive power is 0.24 p.u., the system operates stably. However, at 0.5 s, the reactive power increases to 0.28 p.u., leading to disturbances with weak damping. Prony analysis reveals that the damping is 1.9 rad/s, and the oscillation frequency is identified as 149 Hz. At 1.5 s, when the reactive power rises further to 0.32 p.u., a growing oscillation occurs, causing the instability of the power system. The oscillation damping at 149 Hz is identified as -0.71 p.u., which aligns closely with the theoretically calculated results, thus confirming the accuracy of the analysis results.

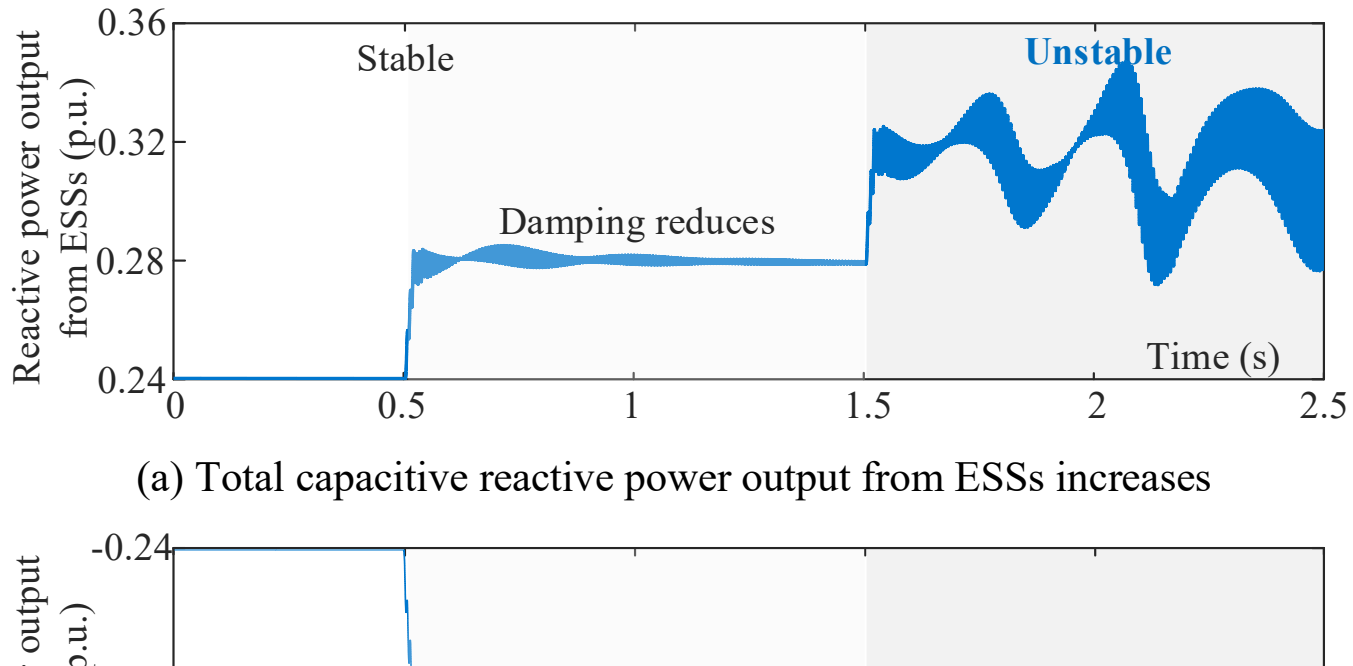

(a) Total capacitive reactive power output from ESSs increases

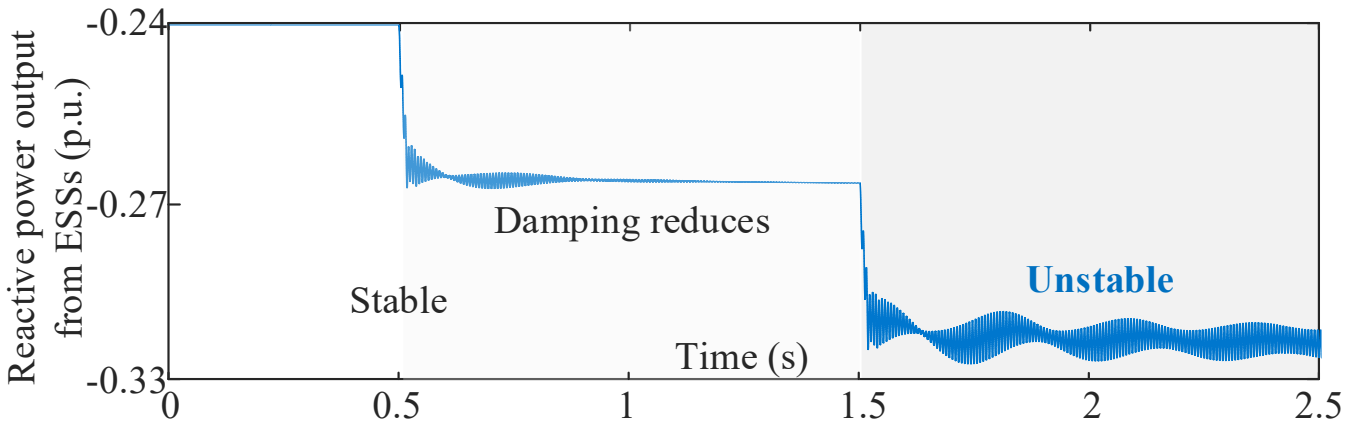

(b) Total inductive reactive power output from ESSs increases

Fig. 14. Simulation results under different operational states

This demonstrates that an excessive increase in the scale of ESSs, particularly in their reactive power output, can negatively impact system stability. Therefore, careful consideration must be given when planning large-scale ESSs in power systems to avoid destabilizing effects on the power system.

*C. Impact of Control Parameter on the Stability*

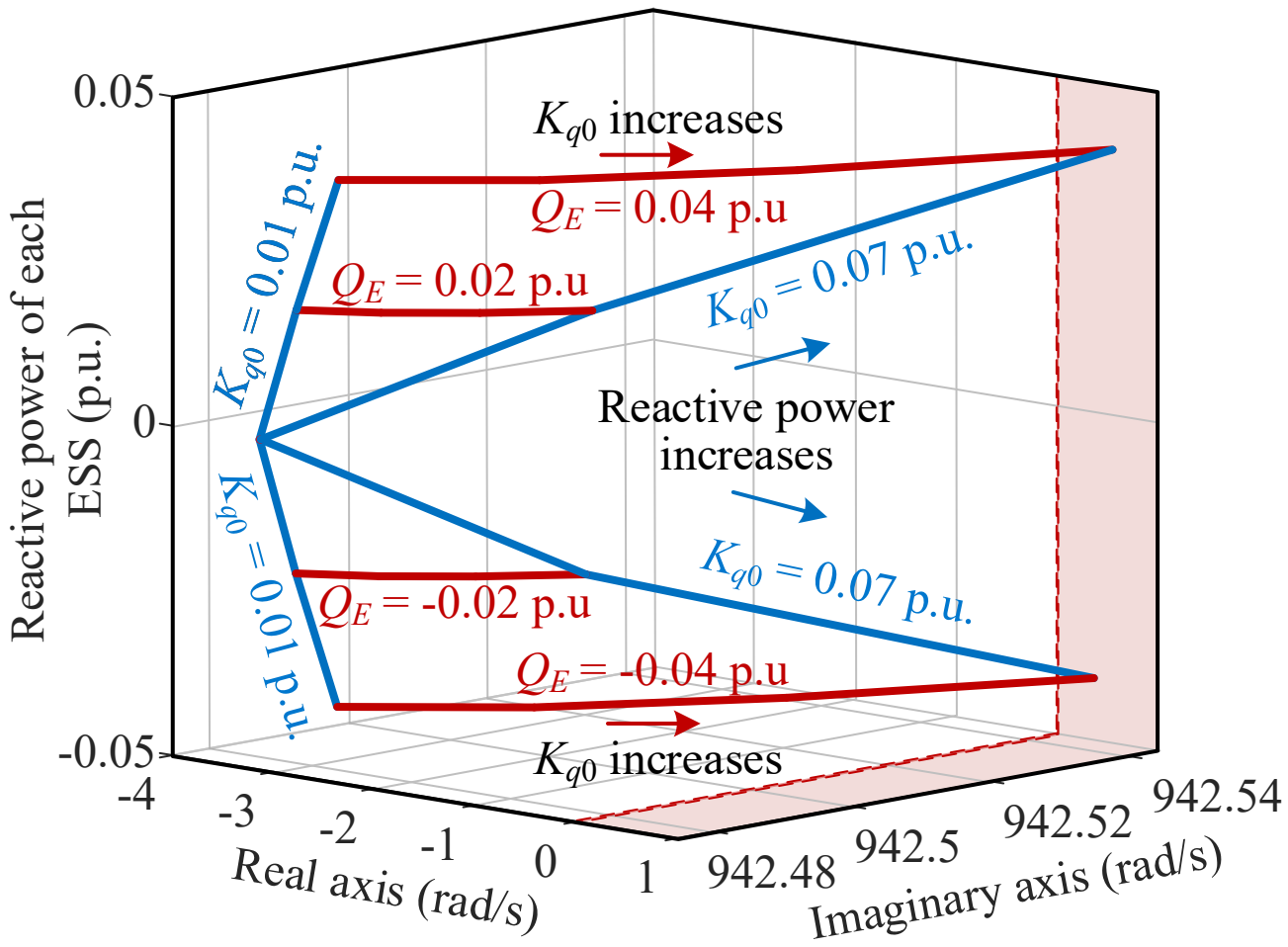

Fig. 15. Trajectories of the oscillation mode as the control parameter changes.

The control parameter $K_{q0}$ affects the stability by changing the dynamic interaction strength among the functional outer control loops of multiple ESSs. Specifically, if $K_{q0}$ = 0, the dynamic interaction disappears, and thus the functional outer control loops do not participate in the dynamic processes of the external power systems.

The impact of $K_{q0}$ under different steady-state operational states is different, as shown by the trajectories in Fig. 15. The oscillation damping decreases as $K_{q0}$ increases, as highlighted by the red lines. For example, the oscillation mode moves from $-2.62 + j942.48$ rad/s to $0.55 + j942.54$ rad/s as $K_{q0}$ increases from 0.01 p.u. to 0.07 p.u., when the total reactive power of ESSs is 0.32 p.u. Similarly, the oscillation mode moves from $-2.62 + j942.48$ rad/s to $0.49 + j942.54$ rad/s when the reactive power is $-0.32$ p.u.

For the same $K_{q0}$, higher reactive power amplitudes result in more significant impacts on oscillation damping, as indicated by the blue lines. For example, the oscillation damping decreases from 3.13 rad/s to 2.62 rad/s as capacitive reactive power increases from 0 p.u. to 0.32 p.u. when $K_{q0}$ = 0.01 p.u. Similarly, the oscillation damping decreases from 3.13 rad/s to -0.55 rad/s when $K_{q0}$ = 0.07 p.u. The results are similar for inductive reactive power.

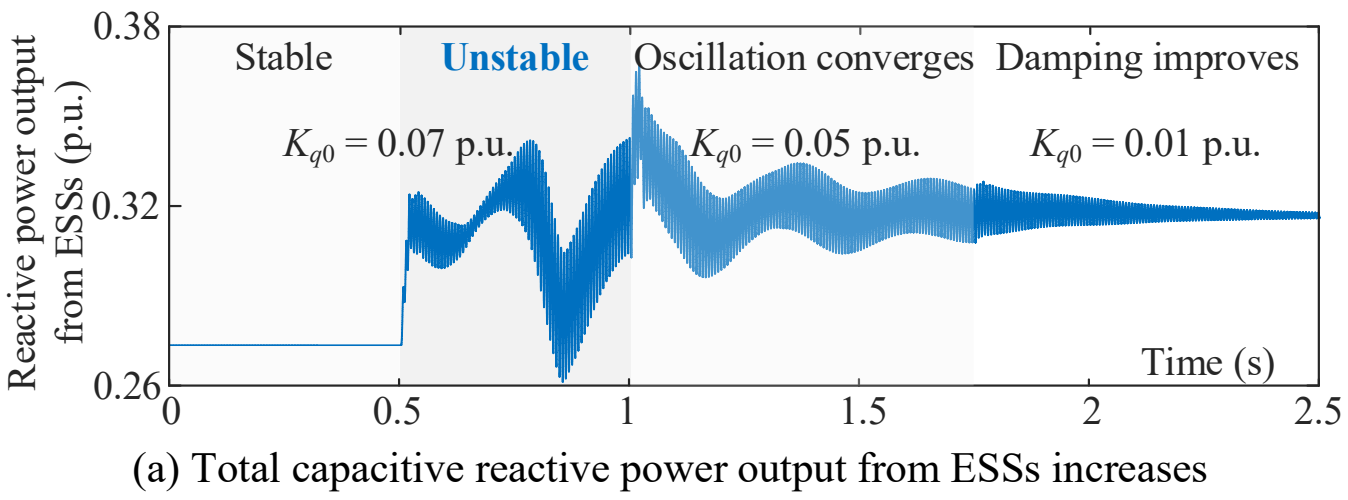

(a) Total capacitive reactive power output from ESSs increases

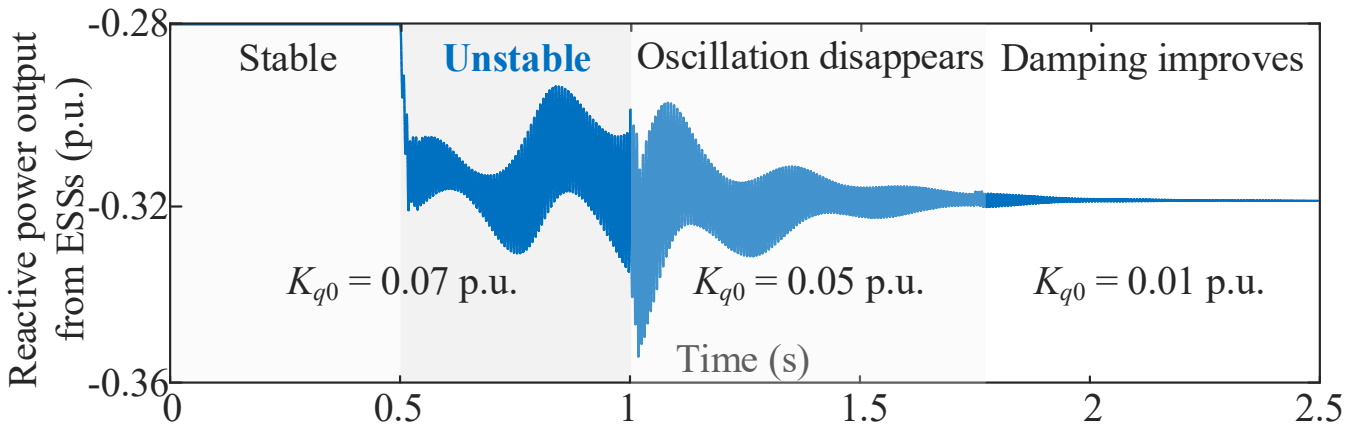

(b) Total inductive reactive power output from ESSs increases

Fig. 16. Simulation results for different control parameter values.

The results in Fig. 16 validate that the oscillation damping decreases as the control parameter $K_{q0}$ increases, regardless of whether the reactive power is inductive or capacitive. However, the system stability improves when $K_{q0}$ is reduced from 0.07 p.u. to 0.05 p.u., and the damping is significantly enhanced when $K_{q0}$ = 0.01 p.u.

To illustrate this more concretely, we focus on the results presented in Fig. 16 (a). Initially, instability occurs when the reactive power of ESSs increases from 0.28 p.u. to 0.32 p.u. At 1 s, the control parameter $K_{q0}$ is reduced to 0.05 p.u., and then the oscillation converges, although with weak damping. Prony analysis shows that the damping is 0.74 rad/s. Subsequently, $K_{q0}$ is further reduced to 0.01 p.u. at 1.75 s, causing a further reduction in the dynamic interaction between the ESSs and the external power system and an improvement in damping. The damping is identified as 2.81 rad/s. This value is close to the theoretical calculation results, confirming the accuracy of the analysis results.

These findings clearly demonstrate that decreasing $K_{q0}$ reduces the dynamic interactions between the ESSs and the external power system, thereby enhancing oscillation damping. Therefore, the results confirm that optimizing $K_{q0}$ is an effective method for improving system stability by mitigating dynamic interactions.

### *D. Impact of the Additional Control Loop on the Stability*

The additional control loop provides extra damping to the power systems, by increasing the value of $K_{dm}$. As $K_{dm} = 0$ p.u., the trajectory of the oscillation mode as the reactive power output from ESSs varies is the same as that in Fig. 13, as highlighted in blue. However, as $K_{dm}$ increases, the oscillation damping improves under all operational states as illustrated in Fig. 17. For example, under the condition that the output reactive power is 0.32 p.u., the damping improves from -0.55 rad/s to 4.17 rad/s when $K_{dm}$ increases from 0 p.u. to 0.07 p.u. and thus, the oscillation is well mitigated, and the power system recovers to be stable. Similarly, the damping increases to 4.22 rad/s from -0.49 rad/s, and to 7.85 rad/s from 3.13 rad/s when the reactive power is -0.32 p.u. and 0 p.u., respectively. It can be well demonstrated that the additional control can provide more damping to the power system and thus enhancing the stability by increasing $K_{dm}$.

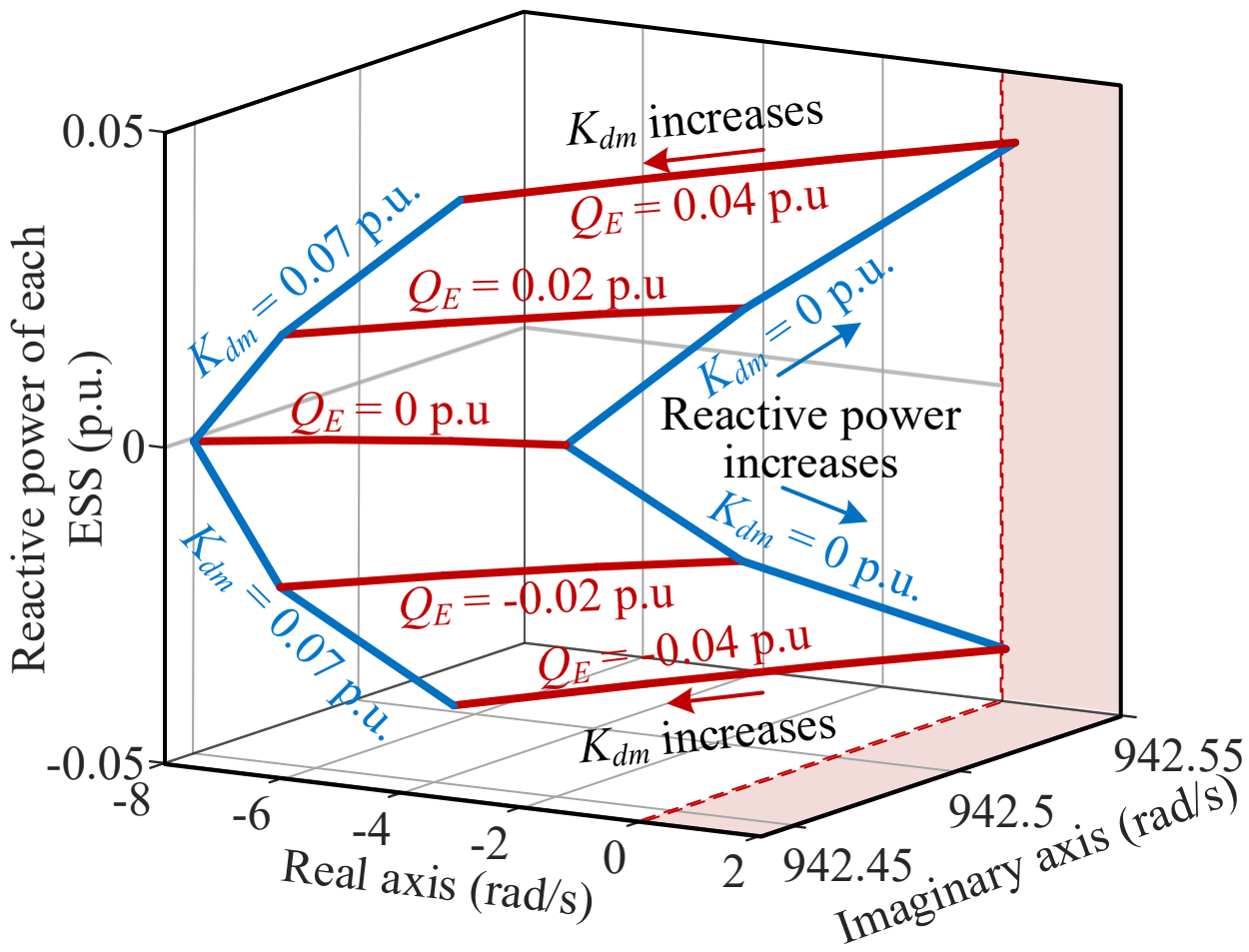

Fig. 17. Trajectories of the oscillation mode with additional control loop.

Moreover, the increased damping is strictly related to $K_{dm}$ instead of operational states, which is different from $K_{q0}$. For example, the damping increased by 4.72 rad/s, 4.71 rad/s, and 4.72 rad/s as $K_{dm}$ increased by 0.07 p.u. under the conditions that the reactive power is 0.4 p.u., 0 p.u., and -0.4 p.u., respectively. Therefore, installing the additional control loop can enhance the stability of the whole power system under any operational condition.

Fig. 18 demonstrates that the oscillation damping improves as the control parameter $K_{dm}$ increases, regardless of whether the reactive power is inductive or capacitive. System stability improves when an additional control loop is installed with a larger $K_{dm}$, and the provided damping becomes notably significant when $K_{dm}$ reaches 0.07 p.u.

To illustrate this in more detail, we focus on the results shown in Fig. 18 (a). Initially, instability occurs at 0.5 s without the additional control loop. After the loop is activated, $K_{dm}$ increases to 0.02 p.u., causing the oscillation to converge, although the damping remains weak. The Prony analysis indicates that the damping is 0.83 rad/s. At 1.75 s, $K_{dm}$ is further increased to 0.07 p.u., leading to a significant improvement in damping. The damping is identified as 4.34 rad/s, which is in close to the theoretical calculation results, confirming the correctness of the findings.

These results clearly show that increasing $K_{dm}$ provides more damping to the oscillation, which, in turn, improves the stability of the system. The study confirms that the addition of the control loop, along with the optimization of $K_{dm}$, is an effective strategy for enhancing oscillation damping and ensuring the stability of the power system.

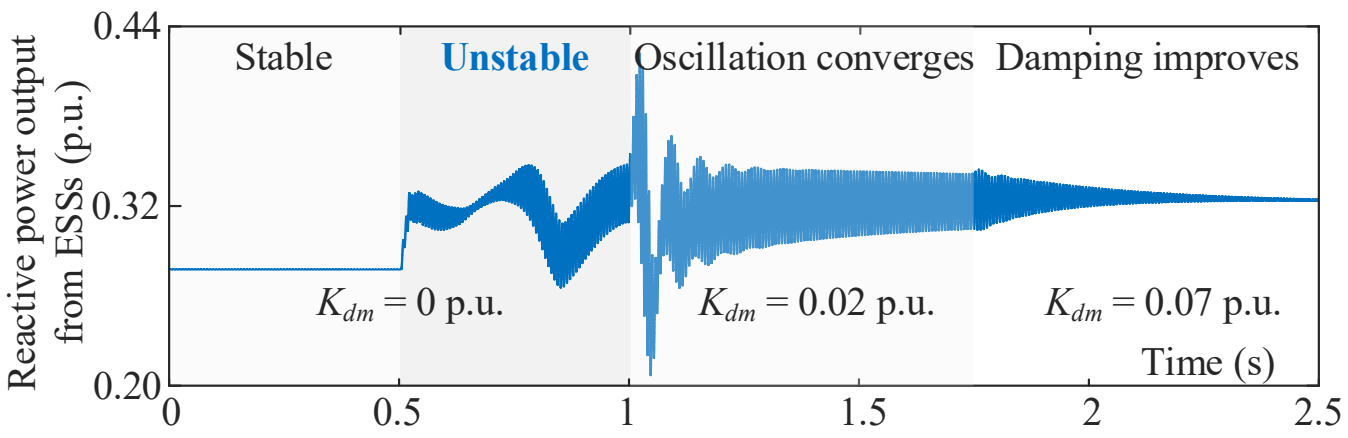

(a) Total capacitive reactive power output from ESSs increases

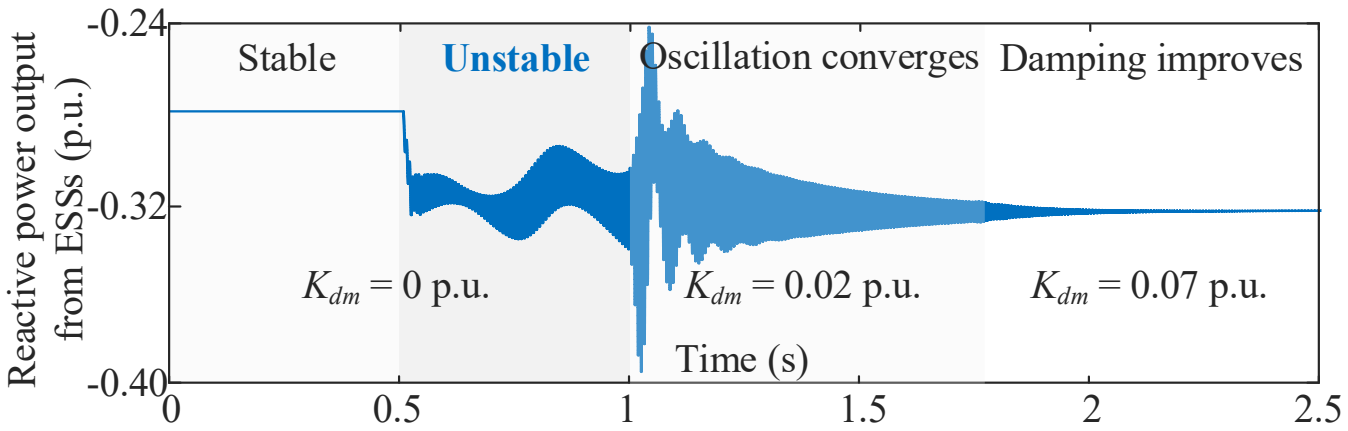

(b) Total inductive reactive power output from ESSs increases

Fig. 18. Simulation results under different operational states

### *E. Impact of Model Errors in ESSs and SGs on the Stability*

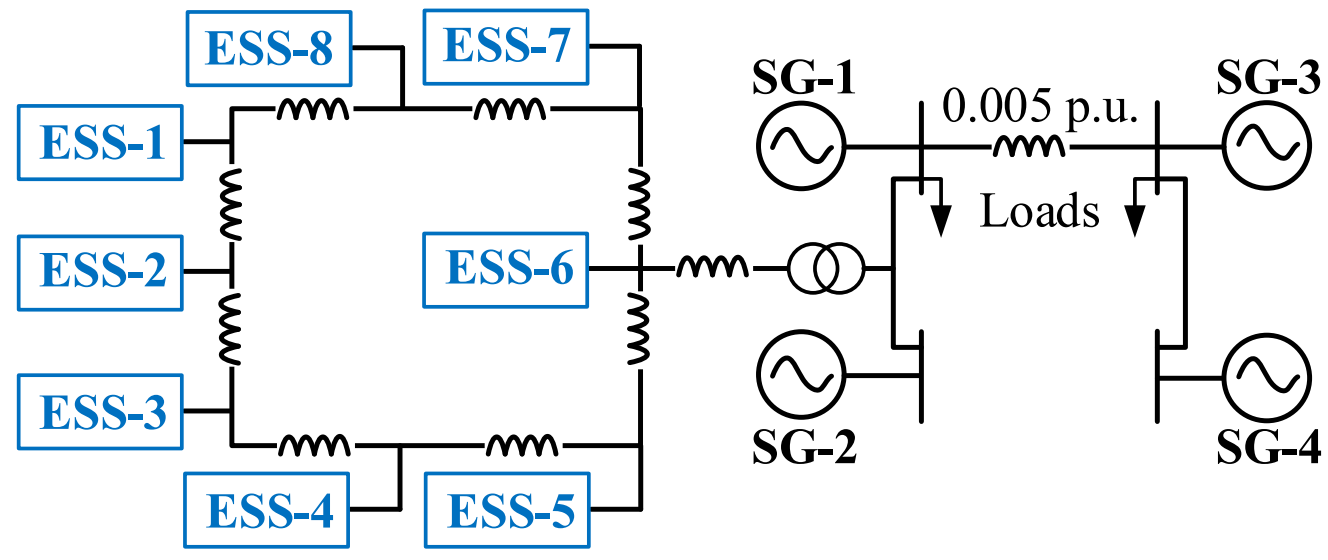

Fig. 19. Configuration of 8 ESSs connected to a four-machine two-area power system.

In Fig. 19, the ESSs are connected to a four-machine two-area power system through a ring network. In the power system, the rated power of each SG is 20 times that of the ESSs, and the AC voltage is 6.3 times that of the ESSs. Thus, the maximum impedance of the transmission line is 0.005 p.u., and the active and reactive power of the loads are 1.6 p.u. and 0.4 p.u., respectively. Initially, all SG inertia constants are set to 2.4 s. The impedance between adjacent ESSs is 0.08 p.u. The reactive power output from ESSs is different and randomly perturbed with a ±10% deviation around the average value.

As the average reactive power of ESSs increases from 0 p.u.

to 0.32 p.u., in both positive and negative power flow directions, the trajectory of the oscillation mode is shown in Fig. 20 with black lines. The moving trend of the oscillation mode is consistent with the observations in previous subsections, that is, as the ESS scale increases, the oscillation damping decreases, which may induce instability. It justifies that the conclusion is also satisfied even when the network topology is changed.

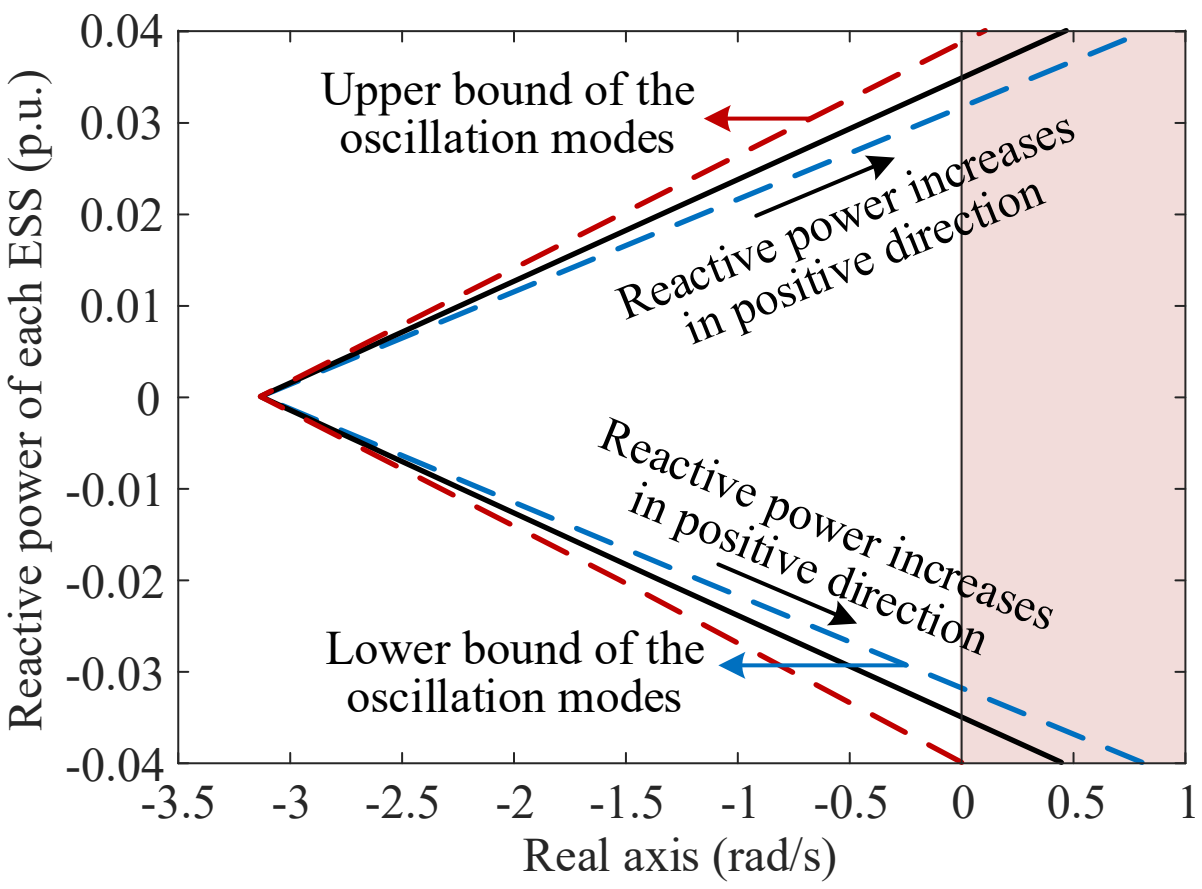

Fig. 20. Trajectories of the oscillation mode with ESS model errors.

However, the precise mode locations vary due to ESS modeling errors. To address this, a sufficient stability criterion under uncertainty of model errors is proposed. Specifically, if the specific values of the model errors are unknown but the up and down boundaries are obtained, the stability can be determined using the boundary values instead of the average values, as shown by the red and blue lines in Fig. 20. For example, with a maximum reactive power error of 10%, the operating point of the critical stability should be changed from 0.32 p.u. to 0.29 p.u. to ensure stability even in the presence of ESS model errors.

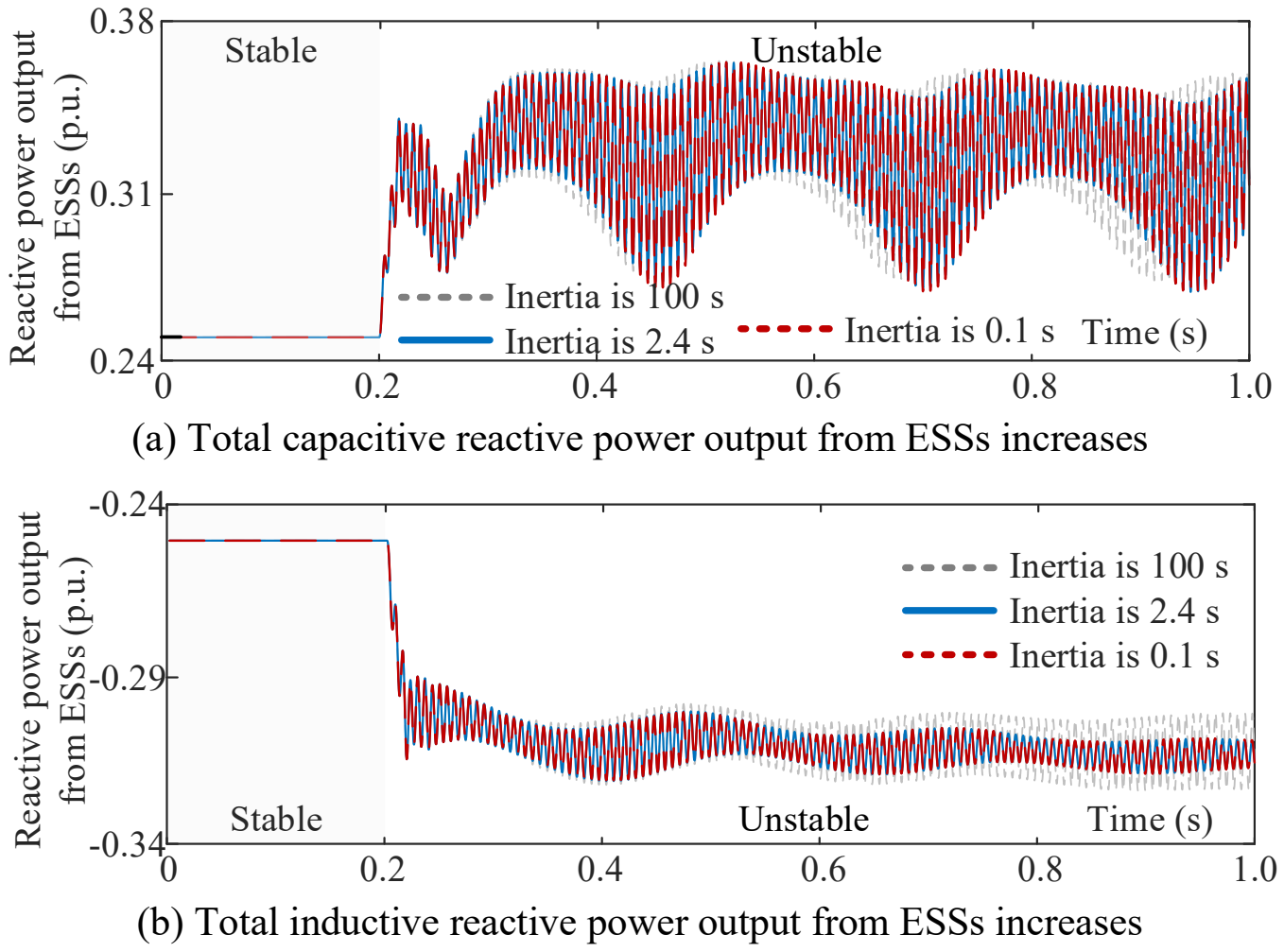

Fig. 21. Simulation results under different inertia of SGs.

In this case study, the SG rated power is 20 times that of the ESSs for a clear observation. In practice, this ratio may be larger. In Fig. 21, the inertia of SGs is changed from 2.4 s to 0.1 s, and 100 s, successively. It can be seen that the ESSs operate stably when the reactive power is 0.25 p.u. As the reactive power increases to 0.32 p.u., the ESSs become unstable. The curves are slightly different under different inertia values. However, the overall impact of this setup remains limited. In practice, the inertia of bulk power systems is typically sufficiently large to maintain the frequency stability. Therefore, modeling the bulk power system as an infinite bus is reasonable for the scenarios considered in this study.

## VII. Conclusions

This paper presents a practical case study of the oscillation occurred under strong grid conditions, identifying the root cause as the superimposed effect of functional outer control loops of ESSs. This finding highlights a critical issue: as the scale of ESS increases, stability challenges become more significant. The functional outer control loops of ESSs provide many benefits, but they also introduce potential stability risks. Furthermore, this study proposes and presents a stability analysis scheme that not only safeguards the confidentiality of ESS manufacturers but also retains sufficient dynamic characteristics of the ESSs for research purposes, and extends the insights from this specific scenario to broader power systems.

The analysis reveals that the oscillation is driven by the superimposed effect of ESSs, which become more pronounced as the scale of ESS expands. It demonstrates that stability risks cannot be ignored simply because of strong grid conditions. More generally, the superimposed effect is more pronounced as the scale of similar equipment increases, not only for ESSs, but also for other IBRs, which is different from the instability caused by the dynamic interactions between two heterogeneous equipment. The key findings and contributions are summarized as follows:

1) ESS functional control loops are widely used to improve power system dynamics and offer significant benefits. However, these control loops also strengthen the dynamic interactions within the clusters of ESSs. These interactions become more significant as the ESS scale increases regardless of the network topology, highlighting a critical stability challenge that cannot be overlooked, even under strong grid conditions.

2) Not all instabilities are directly related to power flow directions, and the oscillation reported in this study is one such example. The oscillation damping is significantly affected by the amplitude of reactive power, regardless of whether it is capacitive or inductive. As the total reactive power amplitude of ESSs increases, the oscillation damping decreases, potentially leading to instability. Therefore, the rated reactive power and number of ESSs should be carefully managed to ensure system stability by further considering the grid conditions.

3) The superimposed effect of ESSs can significantly change the dynamic characteristics of a single ESS, particularly under large-scale conditions. Therefore, an examination involving grid-side dynamics is required, instead of only on the ESS itself. To ensure both the confidentiality of ESS manufacturers and the accuracy of the ESS dynamics, an applicable stability analysis process is proposed and presented in practice. Based on which, the instability risk of large-scale ESSs can be identified and the additional damping control is designed without changing the inner control parameters of ESS PCS.

The superimposed effect caused by large-scale integration of similar equipment is also applicable to IBRs. This inspires both IBR manufacturers and power grid companies to recognize the limitation on the rated power and the number of IBRs in a specific region, even when the grid connection is strong. As the scale of IBRs develops, stability analysis based on an individual IBR is not comprehensive and may overlook risks. Therefore, it is beneficial to evaluate the maximum capacity of IBRs considering system stability limitations.

It is widely observed that information exchange between IBR manufacturers, power grid companies, and research institutes is not fully open-access owing to confidentiality constraints. However, the stability analysis scheme proposed in this paper provides a solution to this issue. Researchers can identify instability risks and propose corresponding solutions from the perspective of the power grid instead of getting involved in the equipment, so the power system can still be ensured to be stable even when control parameters are unknown. This may accelerate the stability analysis process and improve security for large-scale IBR integration.